\documentstyle[epsfig,eqsecnum,floats,preprint,aps]{revtex}
\input epsf
\tighten
\begin{document}
  
\rightline{CU-TP-1118}
\rightline{hep-th/0410142}
\vskip 1cm

\begin{center}
\ \\
\large{{\bf Oscillating bounce solutions and vacuum tunneling in 
de Sitter spacetime}} 
\ \\
\ \\
\ \\
\normalsize{James C. Hackworth\footnote{email address:
hackwort@phys.columbia.edu} and Erick J. Weinberg\footnote{email
address: ejw@phys.columbia.edu} }
\ \\
\ \\
\small{\em Physics Department, Columbia University \\ New York, New
York 10027}

\end{center}

\begin{abstract}

We study a class of oscillating bounce solutions to the Euclidean
field equations for gravity coupled to a scalar field theory with two,
possibly degenerate, vacua.  In these solutions the scalar field
crosses the top of the potential barrier $k>1$ times.  Using analytic
and numerical methods, we examine how the maximum allowed value of $k$
depends on the parameters of the theory.  For a wide class of
potentials $k_{\rm max}$ is determined by the value of the second
derivative of the scalar field potential at the top of the barrier.
However, in other cases, such as potentials with relatively flat
barriers, the determining parameter appears instead to be the value of
this second derivative averaged over the width of the barrier.  As a
byproduct, we gain additional insight into the conditions under which
a Coleman-De Luccia bounce exists.  We discuss the physical
interpretation of these solutions and their implications for vacuum
tunneling transitions in de Sitter spacetime.

\end{abstract}

\setcounter{page}{0}
\thispagestyle{empty}
\maketitle

\section{Introduction}

\begin{figure}[h]
\begin{center}
\epsfig{file=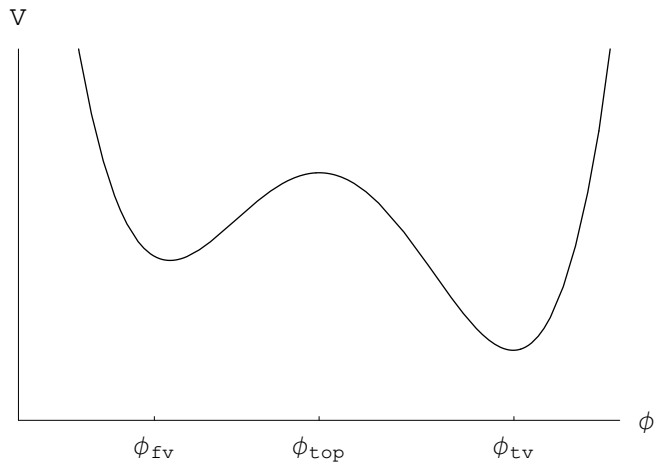,, width =10cm }
\end{center}
\caption{The potential for a typical theory with a false
  vacuum. \label{generic}} 
\end{figure}

There has recently been renewed interest in the problem of vacuum
tunneling in de Sitter spacetime.  Although this interest has been
sparked in large part by developments in string theory, the tunneling
problem itself can be addressed within the context of quantum field
theory.  The essence of the problem is captured by considering a
theory with a single scalar field described by the Lagrangian
\begin{equation}
    {\cal L} = {1 \over 2}(\partial_\mu \phi)^2 - V(\phi)
\label{flatLag}
\end{equation}
where the scalar field potential has two unequal minima, as 
illustrated in Fig.~\ref{generic}.  The lower minimum corresponds to 
the stable ``true vacuum'' state, while the higher minimum
corresponds to a metastable ``false vacuum''.  

At zero temperature, and in the absence of gravitational effects, the
false vacuum decays via a quantum mechanical tunneling process that
leads to the nucleation of bubbles of true vacuum.  The semiclassical
calculation of the bubble nucleation rate per unit volume, $\Gamma$,
is well understood~\cite{ColemanI,ColemanII}.  It can be written in the form
\begin{equation}
  \Gamma = A e^{-B}
\label{GammaForm}
\end{equation}
where $B$ is obtained from the action of the ``bounce'' solution to
the Euclideanized field equations.  This bounce solution has a region
of approximate true vacuum (essentially, a four-dimensional bubble)
separated by a wall region from a false vacuum exterior.

At finite temperature, bubble nucleation proceeds via both quantum
tunneling and thermal fluctuations.  When the temperature $T$ is high
enough that the latter process dominates, $\Gamma$ can still be
written in the form of Eq.~(\ref{GammaForm}), but with $B=E/T$, where
$E$ is the energy of a critical bubble.

In this paper we will be concerned with vacuum tunneling in situations
where gravitational effects are important.  We will assume that
$V(\phi)>0$ everywhere, so that both the false and true vacua
correspond to de Sitter spacetimes.  This problem was first discussed
by Coleman and De Luccia~\cite{ColemanDeLuccia}, who argued that the
proper generalization of the flat spacetime calculation of $B$ could
be obtained by looking for a bounce solution of the Euclidean version
of the coupled matter plus gravity field equations.  When the relevant
mass scales are well below the Planck mass and the bubble sizes are
small compared to the curvature of the de Sitter space, the Coleman-De
Luccia prescription leads to a description of bubble nucleation that
is very similar to that of the zero-temperature flat spacetime case,
with small gravitational corrections of the expected order of
magnitude.

However, when the mass scales are higher or the bubble sizes larger,
not only are the quantitative deviations from the flat space case more
significant, but there are also qualitative differences that raise
issues of interpretation.  These suggest that the de Sitter vacuum
transition process has aspects of both quantum tunneling and of
thermal fluctuation.  Further, the nucleation of true vacuum bubbles
in false vacuum regions seems to imply the possibility of nucleating
false vacuum bubbles in true vacuum regions, with the relative rates
of the two processes having a natural thermal
interpretation~\cite{Lee:1987qc}.

One of the most striking consequences of the inclusion of gravity is
the existence of a homogeneous Euclidean configuration, the
Hawking-Moss solution~\cite{Hawking:1981fz}, that is quite different
in form and physical interpretation from the Coleman-De Luccia bounce,
and that has no counterpart in the flat spacetime problem.

In this article we will focus on yet another type of Euclidean
solution, which might be termed an oscillating bounce.  In contrast
with the Coleman-De Luccia bounce, where the scalar field varies
monotonically from $\phi_{\rm fv}$ toward $\phi_{\rm tv}$, and the
Hawking-Moss solution, where it is everywhere equal to its value at
the top of the barrier, $\phi_{\rm top}$, the field in these solutions
oscillates back and forth between the two sides of the potential
barrier.  Like the Hawking-Moss solution, these have no finite action
counterparts in flat spacetime.  Solutions of this type have also been
discussed by Banks~\cite{Banks:2002nm}.  Here we consider these in
more detail, and examine the conditions under which they can exist.
We find (in disagreement with~\cite{Banks:2002nm}) that, for fixed
values of the parameters of the theory, there are only a finite number
of oscillating bounce solutions.

In Sec.~II, we review the formalism for calculating $\Gamma$ in flat
space-time, both at $T=0$ and at finite temperature.  We emphasize in
particular the aspects that elucidate the physical meaning of the
bounce solution, and discuss the path integral derivation of the
formula for the prefactor $A$ in Eq.~(\ref{GammaForm}).  In Sec.~III
we review how the formalism is generalized to include gravitational
effects, and outline the major features of the Coleman-De Luccia
analysis.  With this preparatory material behind us, we begin our
discussion of new solutions in Sec.~IV.  We develop a framework for
the analysis, and then first apply it to ``small amplitude''
oscillating bounces, for which a linearized approximation can be used.
A critical role is played here by a parameter $\beta$ that measures
the second derivative of the potential at the top of the barrier
relative to the curvature scale of the de Sitter spacetime.  For a
broad class of models, the number of such small amplitude solutions
increases with $\beta$, with new bounces appearing as $\beta$
increases through certain critical values, and with no such solution
at all if $\beta$ is too small.  The bound obtained here is physically
plausible and consistent with previous discussions; however, as we
will see, it is not universally applicable.  In Sec.~V we draw some
intuition from the analysis of the previous section and use it to
discuss bounce solutions, lying outside the small amplitude regime,
for which the effects of the nonlinear terms are dominant.  We then
test this intuition by numerical solution of the bounce equations.  In
Sec.~VI we turn to the case potentials that are unusually flat at the
top, for which the relation between $\beta$ and the number of bounce
solutions that was found in Sec.~IV does not apply.  In the course of
studying these we will gain some insight into the conditions under
which a Coleman-De Luccia bounce can exist.  We will show by explicit
example that bounce solutions can exist even if the potential is
absolutely flat at the top of the barrier, despite some suggestions to
the contrary in the literature.  Finally, in Sec.~VII we discuss the
physical interpretation of the oscillation bounce solutions and
include some concluding remarks.

\section{Review of vacuum decay in flat spacetime}
\subsection{Flat space tunneling at zero temperature}

We first recall some results concerning quantum tunneling in a system 
with more than one degree of freedom.  Consider a 
system with coordinates $q^j$ ($j=1,\dots,N$) whose dynamics is 
governed by the Lagrangian
\begin{equation}
   L = {m\over 2} \left({dq^j\over dt}\right)^2  - U(q)
\end{equation}
and let the point $q_{\rm fv}$ be a local (but not global) minimum of the
potential energy.  Given a system whose wave function is initially
localized about $q_{\rm fv}$, we want to know the rate at which the system
tunnels through the surrounding potential barrier.

For each path $q(s)$ through the potential barrier that begins at
$q(s_1) = q_{\rm fv}$ and emerges from the barrier at a
point $q(s_2) \equiv q_2$, one can calculate a one-dimensional WKB
tunneling integral
\begin{equation}
      B = 2 \int_{s_1}^{s_2} ds \sqrt{2m[U(q(s))- U_{\rm fv}]} \, \, ,
\label{barrierintegral}
\end{equation}
where $U_{\rm fv} \equiv U(q_{\rm fv})= U(q_2)$.  The leading WKB
approximation to the tunneling rate is proportional to $e^{-B}$,
evaluated along the path that minimizes the tunneling
integral~\cite{Banks:1973ps}.  The end point $q_2$ of this path is the
most probable place for the system to emerge from the barrier, and
thus gives the initial condition for the classical evolution of the
system after tunneling.

By manipulations analogous to ones familiar from classical mechanics,
the problem of minimizing the integral in
Eq.~(\ref{barrierintegral}) can be recast as the problem of finding a
stationary point of the Euclidean action
\begin{equation}
   S_E =  \int_{\tau_1}^{\tau_2} d\tau \left[ {m\over 2}
      \left({dq^j\over d\tau}\right)^2   + U(q)\right] \, .
\end{equation}
The solution $q_{\rm b}(\tau)$ of the Euclidean Euler-Lagrange
equations gives the same path as before, although 
with a different parameterization.  Because
$q_{\rm fv}$ is a minimum of $U$, the initial Euclidean time must be taken
as $\tau_1 = -\infty$; the final time $\tau_2$ is arbitrary.

Since $dq_{\rm b}/d\tau$ vanishes at $\tau_2$, this solution can be
continued, in a ``$\tau$-reversed'' form, to give a ``bounce''
solution that runs from $q_{\rm fv}$ to $q_2$ and then back to
$q_{\rm fv}$.  The exponent in the tunneling factor is then
\begin{eqnarray}
   B &=& \int^\infty_{-\infty} d\tau \left[ {m\over 2}
      \left({dq^j_{\rm b}\over d\tau}\right)^2 + U(q_{\rm b}) -
      U(q_{\rm fv})\right] \cr  \cr
       &=& S_E(q_{\rm b}) - S_E(q_{\rm fv})  \, ,
\end{eqnarray}
where the factor of 2 in Eq.~(\ref{barrierintegral}) has been absorbed
by the doubling of the action that results from considering the full
bounce.

Adapting~\cite{ColemanI} this formalism to the scalar field theory of
Eq.~(\ref{flatLag}), one is led to consider the Euclidean action
\begin{eqnarray}
   S_E &=&  \int dx_4 \, d^3{\bf x} \left[ 
     {1\over 2} \left({\partial \phi\over \partial x_4}\right)^2
     +{1\over 2} ({\bf \nabla}\phi)^2 + V(\phi) \right]    \cr  \cr \cr
      &=&  \int d^4x \left[ {1\over 2} (\partial_a \phi)^2
           +  V(\phi) \right]
\end{eqnarray}
and to seek a bounce solution of the Euclidean field equation
\begin{equation}
    \partial_a \partial_a \phi = {dV\over d\phi}  \, .
\label{fullEuclideanequation}
\end{equation}
To match the initial state before tunneling, this solution must
tend to $\phi_{\rm fv}$ as $x_4 \rightarrow \pm\infty$, while the 
finiteness of the tunneling exponent
\begin{equation}
    B = S_E(\phi_{\rm bounce}) - S_E(\phi_{\rm fv})
\label{fieldBfactor}
\end{equation}
requires that it also tend to $\phi_{\rm fv}$ at spatial infinity.
The bounce contains an interior region, in which
the field is on the true vacuum side of the barrier, that is separated
from the false vacuum exterior by a wall of finite thickness.  A
spatial slice through the center of this solution gives the
three-dimensional configuration that is both the optimal end point of
the quantum tunneling and the initial condition for the subsequent
classical evolution.  This configuration contains a true vacuum bubble
embedded in the false vacuum, with the total potential energy (i.e.,
the sum of the gradient energy and the scalar field potential) being
equal to the initial false vacuum energy.

This WKB calculation only gives the exponential factor in the
tunneling rate.  The pre-exponential factor is most easily obtained by
path integral methods~\cite{ColemanII}.  The basic strategy is to use
a Euclidean path 
integral to calculate the energy of the false vacuum state.  Since
this is an unstable state, its ``energy'' is complex, with its
imaginary part giving the decay rate.

Specifically, consider the path integral
\begin{equation}
    I({\cal T}) = \int [d\phi] \, e^{-S_E(\phi)} \, ,
\end{equation}
where the integration is restricted to paths obeying $\phi({\bf x},
-{\cal T}/2) = \phi({\bf x},{\cal T}/2) = \phi_{\rm fv}$.  As ${\cal T}
\rightarrow \infty$ the path integral is dominated by the lowest
energy state with nontrivial overlap with these boundary conditions;
i.e., by the false vacuum.  Hence
\begin{equation}
     E_{\rm fv} = - \lim_{\cal T \rightarrow \infty} 
         { \ln I({\cal T}) \over \cal T}
\label{EfromPath}
\end{equation}
and the bubble nucleation rate per unit volume is 
\begin{equation}
    \Gamma = 2 \lim_{{\cal T},\, \Omega \rightarrow \infty} 
      \left[{ {\rm Im}\, \ln I({\cal T}) \over \Omega \cal T}
           \right] \, ,
\label{GammaFromPath}
\end{equation}
where $\Omega$ is the volume of space, assumed to be taken to
infinity at the end of the calculation.

The path integral can be evaluated by summing the contributions from
its stationary points.  The first of these is simply a constant
homogeneous false vacuum configuration, $\phi(x) = \phi_{\rm fv}$.  To
leading order, the contribution of this to the path integral is
\begin{equation}
    I_0({\cal T}) =  e^{-S_E(\phi_{\rm fv})} \,  
    \left[\det S''_E(\phi_{\rm fv})\right]^{-1/2}   \, ,
\end{equation}
with $S''_E(\phi_{\rm fv})$ denoting the second variation of the
action about the (constant) classical solution.

Next is the bounce solution, $\phi_{\rm b}(x)$.  In calculating the
determinant factor here, one finds that $S''_E(\phi_{\rm b})$ has
four\footnote{In theories with internal symmetries there may be
additional zero modes; for an example of this, see
Ref.~\cite{KusenkoLeeWeinberg}.} zero modes and one negative mode.
The former must be replaced by collective coordinates specifying the
location in space and Euclidean time of the bounce, while the latter
gives a factor of $i$ when the square root of the determinant is
taken.  The contribution to the path integral is
\begin{equation}
    I_1({\cal T}) = {i \over 2} \Omega {\cal T} J \,
           \left|\det{}'S''_E(\phi_{\rm b})\right|^{-1/2} \,
           e^{-S_E(\phi_{b})} \equiv {i\over 2} \Omega {\cal T} 
          K\, e^{-B} \, I_0({\cal T}) \, ,
\end{equation}
where $J$ is the Jacobean factor from replacing the zero modes
by collective coordinates, the factor of $\Omega {\cal T}$ is
from the integration over the collective coordinates, the 1/2 comes
from a careful treatment of the negative mode integration, and the
prime on the determinant indicates that it does not include the zero
modes.

Finally, there are the contributions from the approximate stationary
points corresponding to many widely separated bounces.\footnote{The
dilute gas approximation of considering only widely separated bounces
is valid for $B\gg 1$.}  These are of the form
\begin{equation}
    I_n({\cal T}) 
       = {1 \over n!} \left[{i\over 2} \Omega {\cal T} K\right]^n 
                 \,  e^{-nB} \, I_0({\cal T})  \, ,
\end{equation}
where the factorial enters because interchanging bounces does not give
a new contribution.  Summing over $n$ gives
\begin{equation}
    I({\cal T}) = \sum_{n=0}^\infty I_n({\cal T}) = I_0({\cal T})
         \exp\left[ {i\over 2} \Omega {\cal T} K e^{-B} \right]  \, .
\end{equation}
The factor of $I_0({\cal T})$ only contributes to the real part of the
energy, so Eq.~(\ref{GammaFromPath}) gives
\begin{equation}
    \Gamma = K e^{-B}  \, .
\end{equation}

The bounce equation is usually solved by assuming O(4) symmetry, so
that $\phi$ is a function only of $s =\sqrt{{\bf x}^2 + x_4^2}$.  
Equation~(\ref{fullEuclideanequation}) then reduces to 
\begin{equation}
    {d^2 \phi \over ds^2} + {3 \over s}\, {d \phi \over ds}
          = {dV\over d\phi}
\label{flatPhiEq}
\end{equation}
with the boundary conditions 
\begin{equation}
     \left.{d \phi \over ds}\right|_{s=0} =0  \, , 
        \qquad \phi(\infty)=\phi_{\rm fv}   \, .
\label{flatBdy}
\end{equation}

The analysis of this equation is aided by considering an analogous
problem in which $s$ is time and $\phi$ represents the position of a
unit mass particle with potential energy $U=-V$ that is subject to a
frictional force proportional to $3/s$. Finding a solution is
tantamount to finding an initial ``position'' $\phi(0)$ such that as
$s \rightarrow \infty$ the particle comes to rest at $\phi_{\rm fv}$,
where $U=-V$ has a local maximum.  The existence of such a position
can be established by Coleman's ``overshoot/undershoot''
argument~\cite{ColemanI}, which relies on the fact that the friction
decreases monotonically with time.

Several points should be noted:

1) The fact that the fluctuations about the bounce included only a
   single negative mode played an essential role in the path integral
   derivation.  The false vacuum energy would have been real if there
   had been an even number of negative modes, while for an odd number
   of the form $4k+3$, the imaginary part of the energy would have had
   the wrong sign. 

2) Although the O(4) symmetry of the Euclidean Lagrangian is
   technically quite useful, it does hide a very real asymmetry in the
   physical interpretation of the $x_a$.  While $x_1$, $x_2$, and
   $x_3$ are ordinary spatial variables, $x_4$ is simply a convenient
   parameterization of the configurations $\phi({\bf x}; x_4)$ that
   define the optimal tunneling path.

3) Although the one-bounce solution has a natural interpretation in
   terms of the optimal tunneling path, there seems to be no analogous
   simple interpretation for the multi-bounce solutions.

\subsection{Flat space tunneling at finite temperature}

This formalism can be readily
extended~\cite{Langer:1969bc,Linde:1981zj} to the case of finite
temperature $T$.  Instead of obtaining the bubble nucleation rate in
terms of the imaginary part of the energy of the false vacuum, one
instead calculates it in terms of the imaginary part of the free
energy of the false vacuum.  This can be obtained by recalling that
the partition function is given by a path integral over configurations
that are periodic in Euclidean time with period $1/T$.  Hence, the
bounce solutions satisfy the Euclidean field equations on $R^3 \times
S^1$ rather than on $R^4$.  The only boundary conditions are at
spatial infinity, where $\phi$ is required to take on its false vacuum
value.

Two quite different types of periodic bounce solutions immediately
come to mind.  At low temperatures (i.e., when $1/T$ is much greater
than the characteristic bounce radius), a periodic bounce can be
obtained by a small deformation of the zero temperature bounce.
The action of this solution, which clearly corresponds to quantum
tunneling, differs only slightly from that when $T=0$.

In the second type of bounce, corresponding to bubble nucleation by
thermal fluctuations, the fields are independent of Euclidean time and
so trivially satisfy the periodicity conditions.  Since the integral
over the Euclidean time is trivial, the action can be written as $S =
E/T$, where $E$ is the free energy of the static solution of the
three-dimensional field equations.\footnote{To be more precise, $E$ is
actually a temperature-dependent quantity $E(T)$.  In the path
integral formalism this temperature-dependence comes about from a
reordering of counterterms in the Lagrangian, corresponding to the
standard field theory calculations~\cite{Dolan:1973qd,Weinberg:1974hy}
of the high temperature effective potential.}  These thermal bounces
are subdominant at low temperature, but dominate at high temperature.

It is instructive to contrast the spatial slices of the vacuum
tunneling bounce with those of the thermal bounce.  In the former
case, the hypersurface at $x_4 = -\infty$ gives the initial false
vacuum configuration before tunneling, and a slice through the center
of the bounce gives the optimal exit point from the potential energy
barrier, which is a configuration containing a supercritical bubble
that will expand from rest.  In the latter case, the spatial
hypersurfaces give a static configuration containing a critical
bubble, whose radius is at the top of the potential energy barrier,
balanced between expansion and contraction.  The initial state is only
evident from the asymptotic behavior at spatial infinity.

\section{Vacuum tunneling in curved spacetime}

Coleman and De Luccia~\cite{ColemanDeLuccia} extended the bounce
formalism to include the gravitational effects on vacuum decay.  They
argued that one should an Einstein-Hilbert term to the Euclidean
action and then seek bounce solutions of the resulting Euclidean field
equations.  The tunneling exponent would then again be given by
Eq.~(\ref{fieldBfactor}), but with the actions now including the
additional gravitational terms.  Their discussion did not, however,
include the calculation of the prefactor $A$, an issue that remains
poorly understood.

If one assumes O(4) spherical symmetry, the metric can be written in
the form
\begin{equation}
    ds^2 = d\xi^2 + \rho(\xi)^2 d\Omega_3^2   \, , 
\label{SpherSymMetric}
\end{equation}
where $d\Omega_3^2$ is the metric on the unit three-sphere, and 
all scalar fields depend only on $\xi$.
The action reduces to 
\begin{equation}
    S_E = 2\pi^2 \int d\xi \left[\rho^3 
        \left({1\over 2} \dot\phi^2 + V \right)
   + {3M_{\rm Pl}^2 \over 8\pi} 
        \left( \rho^2 \,\ddot \rho + \rho\,\dot\rho^2 - \rho \right) \right]
\label{curvedaction}
\end{equation}
with overdots denoting differentiation with respect to $\xi$.

The Euler-Lagrange equations are summarized by
\begin{equation}
    \ddot \phi + {3 \dot\rho \over \rho} \, \dot\phi = {dV\over d\phi}
\label{ddotPhiEq}
\end{equation}
and 
\begin{equation}
    {\dot \rho}^2 = 1 + {8 \pi \over 3 M_{Pl}^2}\,\rho^2
          \,\left({1\over 2} \dot\phi^2 - V \right)   \, .
\label{EucFriedmann}
\end{equation}
Because of the second term in Eq.~(\ref{ddotPhiEq}), $\dot\phi$ must
vanish at the zeros of $\rho$.  If $V(\phi) \ge 0$, which we will
assume, there are always two such zeroes, so that the Euclidean
manifold is topologically a four-sphere~\cite{Guth:1982pn}; the only
exception occurs if $V(\phi)$ vanishes at one of its minima, in which
case the homogeneous solution with $\phi$ at this minimum has the flat
metric of $R^4$.  One of the zeroes of $\rho$ can be taken to be
$\xi=0$.  If the second zero is denoted $\xi_{\rm max}$, then the
boundary conditions on $\phi$ are that
\begin{equation}
    \dot \phi(0) =  \dot\phi(\xi_{\rm max}) = 0 \, .
\label{curvedBdy}
\end{equation}
The symmetry of these boundary conditions is in contrast with 
the flat space boundary conditions of Eq.~(\ref{flatBdy}).

One consequence of the four-sphere topology is that the Euclidean
actions of the bounce and the false vacuum are separately finite.  In
flat space, $S_E(\phi_{\rm fv})$ is formally infinite, and
Eq.~(\ref{fieldBfactor}) only makes sense in terms of a point-by-point
subtraction of $V(\phi_{\rm fv})$ inside the spatial integral.  In
contrast, the Euclidean false vacuum solution of
Eqs.~(\ref{ddotPhiEq}) and (\ref{EucFriedmann}) is a four-sphere of
radius $H_{\rm f}^{-1}$, with
\begin{equation}
     \rho(\xi) = H_{\rm f}^{-1} \sin (H_{\rm f} \xi)  
\label{fvRho}
\end{equation}
and 
\begin{equation}
     H_{\rm f} \equiv  \sqrt{8 \pi V(\phi_{\rm fv})
                \over3 M_{\rm Pl}^2} \, .
\label{fvHubble}
\end{equation}
Its action is
\begin{equation}
    S_E(\phi_{\rm fv}) = - {3 \over 8}\, {M_{\rm Pl}^4 \over
           V(\phi_{\rm fv}) }   \, ,
\label{fvAction}
\end{equation}
with the factor of $-3/8$ being the sum of $+3/8$ from the matter term
in Eq.~(\ref{curvedaction}) and $-3/4$ from the second, gravity, term.

One would expect gravitational effects on tunneling to be small if the
characteristic scales of $V$ are much less than the Planck mass, and
the flat space bounce radius $\bar r \ll H_{\rm f}^{-1}$.  In this
regime, the Coleman-De Luccia bounce solution has a central true
vacuum region, $0 \le \xi \lesssim \bar r$, where the scalar field
profile closely approximates that of the flat space bounce.  Outside
this region $\phi$ rapidly approaches its false vacuum value although,
because $\xi_{\rm max}$ is finite, $\phi_{\rm b}(\xi) - \phi_{\rm fv}$
never quite vanishes.  The tunneling exponent $B$ differs from the
flat space value by a fractional amount of order $ \bar r^2H_{\rm
f}^2$.

\begin{figure}[bt]
\begin{center}
\epsfig{file=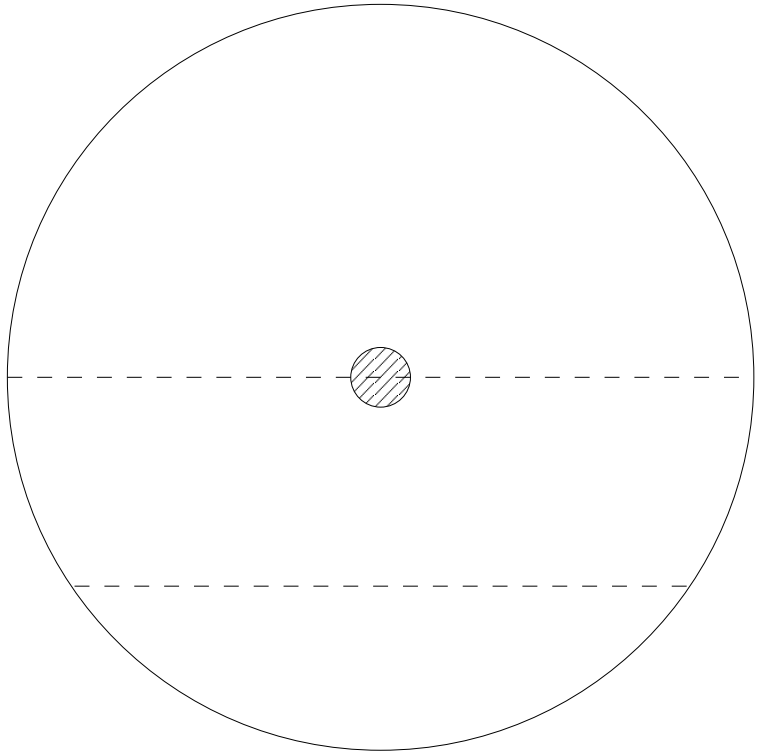, width =10cm }
\end{center}
\caption{ A Coleman-De Luccia bounce solution for the case where the
flat space bounce radius is much less than $H_{\rm f}^{-1}$.  The
picture should be visualized as a four-sphere viewed head-on, with
$\xi=0$ being the point at the center of the cross-hatched region
where the field is on the true vacuum side of the potential barrier;
$\xi=\xi_{\rm max}$ is the antipodal point on the opposite side of the
sphere.  The dashed line passing through $\xi=0$ (and also through
$\xi=\xi_{\rm max}$) denotes a three-sphere corresponding to the
spatial hypersurface on which the bubble materializes.  The
three-sphere denoted by the lower dashed line is roughly analogous to
the false vacuum initial-state hypersurface in the flat space problem.
\label{foursphere}}
\end{figure}

In this regime, the physical interpretation of the bounce solution
carries over with only slight modifications from the flat space case.
As before, a three-dimensional slice through the center of the bounce
(see Fig.~\ref{foursphere}) can be viewed as giving initial data for
the classical evolution; the fact that this slice has finite volume
reflects the finite volume of (most) spacelike slices through de
Sitter spacetime.  In flat space, the hypersurface at $x_4 = -
\infty$, with $\phi(x)$ identically equal to its false vacuum value,
represented the initial state before tunneling.  Although this has no
precise counterpart here, one can view a slice such as that shown by
the lower dashed line in Fig.~\ref{foursphere} as being analogous to a slice
at large negative $x_4$ in the flat case.  There are also multibounce
configurations that are approximate stationary points of the action,
just as in the flat space case.  Summing the contributions of these to
the Euclidean path integral, one would obtain an exponential of the
one-bounce contribution (modulo the difficulties of properly defining
the fluctuation determinant), just as in the previous case.

There is, however, one difference, related to the point $\xi=\xi_{\rm
max}$, that should be pointed out here.  One might have expected this
to be analogous to $s=\infty$ in the flat space bounce.  In the
latter, $s=\infty$ defines a three-sphere and corresponds both to
spatial infinity on the hypersurface where the bubble materializes and
to the initial state hypersurface.  By contrast, $\xi=\xi_{\rm max}$
is a single point that lies on the spacelike hypersurface that gives
initial data for the Lorentzian problem.  However, as can be seen from
Fig.~\ref{foursphere}, it does not lie on the initial state
hypersurface.

The deviations from the flat space case become quite significant when
the flat space bounce radius is comparable to or greater than
$H_f^{-1}$.  In some such cases, there is no Coleman-De Luccia bounce
at all.  When there is a bounce, its true vacuum region (whose size
may be very different than in the flat space case) occupies a
significant fraction of the Euclidean four-sphere, and one can no
longer identify even an approximate initial-state hypersurface.
Indeed, the asymmetry between the true vacuum interior and the false
vacuum exterior is lost to a large degree.  This suggests that the two
regions can be viewed as being on equal footings, and that the bounce
can describe either the production of a true vacuum bubble in a false
vacuum background or that of a false vacuum bubble in a true vacuum
background, with the initial state determined not by the properties of
the bounce solution, but by whether $S_E(\phi_{\rm fv})$ or
$S_E(\phi_{\rm tv})$ is subtracted when calculating $B$.  (A detailed
examination of this possibility shows that the rates for nucleation of
false vacuum bubbles in true vacuum and true vacuum bubbles in false
vacuum are related by a factor that has a natural thermal
interpretation, and that takes a simple Boltzmann form in some
limits~\cite{Lee:1987qc}.)

Finally, note that any multibounce solutions are clearly limited and quite
constrained in this large-bounce regime, so there is no exponentiation of the
bounce factor; this fact by itself gives a hint of the difficulties of
generalizing the path integral calculation of $\Gamma$.

Whether or not there is a Coleman-De Luccia bounce solution, there is
always a second type of solution, first pointed out by Hawking and
Moss~\cite{Hawking:1981fz}, that has no analogue in flat spacetime.
Like the pure false vacuum solution, this is homogeneous, but with the
scalar field everywhere equal to $\phi_{\rm top}$, its value at the
top of the barrier.  Calculations exactly analogous to those leading
to Eqs.~(\ref{fvRho})-(\ref{fvAction}) give
\begin{equation}
     \rho(\xi) = H_{\rm top}^{-1} \sin (H_{\rm top} \xi)  
\end{equation}
and
\begin{equation}
   B_{\rm HM} = S_E(\phi_{\rm top}) - S_E(\phi_{\rm fv}) = 
       {3 \over 8}\, \left [-{M_{\rm Pl}^4 \over
           V(\phi_{\rm top}) } + {M_{\rm Pl}^4 \over
           V(\phi_{\rm fv})} \right] \, .
\end{equation}
with $H_{\rm top}$ defined by a formula analogous to
Eq.~(\ref{fvHubble}).

\section{new solutions}

Our focus in this paper is on a third class of Euclidean solutions
that are neither simple bounces nor Hawking-Moss.  We continue to
assume O(4) symmetry, so Eqs.~(\ref{SpherSymMetric})-(\ref{curvedBdy})
still apply.  However, we now allow $\phi$ to cross the barrier an
arbitrary number of times between $\xi=0$ and $\xi_{\rm max}$; we will
denote this number by $k$, so that $k=1$ corresponds to the Coleman-De
Luccia bounce.  The possibility of multiple barrier crossings is in
contrast with the flat space case, where only $k=1$ solutions can
exist.

To simplify matters, we will restrict ourselves to the case of a
potential with only two local minima.  Without any loss of generality
we can assume that $\phi_{\rm fv} < \phi_{\rm top} < \phi_{\rm tv}$.

Our problem can be viewed as that of finding values of $\phi_0$ for
which the initial conditions $\phi(0) = \phi_0$, $\dot\phi(0) =
\rho(0) =0$ imply that $\dot\phi(\xi_{\rm max}) = 0$.  By scanning the
range $\phi_{\rm fv} < \phi < \phi_{\rm tv}$ for such critical values,
all bounce solutions can be found.  In fact, because the problem is
invariant under the transformation $\xi \rightarrow \xi_{\rm max} -
\xi$, every solution with odd $k$ will be encountered twice in such a
scan, once as a ``false-to-true'' bounce and once as a
``true-to-false'' one.  There may also be false-to-false and
true-to-true solutions for which the initial and final values of
$\phi$ are on the same side of the barrier.  These appear either once
or twice during the scan, depending on whether or not the two
endpoints are identical; we have found examples of both types of
behavior.

It should be stressed that our use here of phrases such as
``false-to-true'' refers only to the variation of $\phi$ as $\xi$
ranges from 0 to $\xi_{\rm max}$.  As was pointed out in the previous
section, the value of the field at $\xi_{\rm max}$ [in contrast to
$\phi(\infty)$ in the flat space bounce] is not necessarily related to
the initial state.

In the flat space problem, the possible values of $\phi_0$ can be divided
into undershoot and overshoot regions, with the critical
bounce value lying at the boundary between the regions.  In the curved
space case, where trajectories can cross $\phi_{\rm top}$ more than
once, this overshoot/undershoot classification becomes somewhat
ambiguous.  Instead, we find it more useful to define a function
$h(\phi_0)$ whose zeroes correspond to the critical trajectories.

An obvious candidate for this function would be the value of
$\dot\phi(\xi_{\rm max})$ on the trajectory specified by $\phi_0$.
However, this diverges whenever it is nonzero, and so does not give a
satisfactory $h(\phi_0)$.  The quantity $\rho^3 \dot\phi$ is better
behaved, but it too can diverge if the trajectory goes into a region
where $dV/d\phi$ is unbounded.  However, we can avoid any such
divergence by modifying the growth of $V$ at large $\phi$.  This will
have no effect on the critical trajectories, since these remain within
the finite interval $\phi_{\rm fv} < \phi < \phi_{\rm tv}$.

Let us therefore define $h(\phi_0)$ to be the value of $\rho^3
\dot\phi$ at $\xi_{\rm max}$, evaluated on the trajectories of the
modified potential.  The zeros of $h$ are the values of $\phi_0$ that
give bounce solutions satisfying the boundary conditions.  Although
$h$ depends in detail upon precisely how the potential was modified,
the locations and nature of its zeros do not, and it is really only
these with which we are concerned.  There will always be zeros at
$\phi_{\rm fv}$, $\phi_{\rm tv}$, and $\phi_{\rm top}$, corresponding
to the pure false vacuum, pure true vacuum, and Hawking-Moss
solutions, respectively.  Depending on the potential these may be the
only zeros of $h$, or there may be others.

The plot of $h(\phi_0)$ evolves as the parameters of the theory are
varied.  During the course of this evolution the locations of the
zeros will vary continuously.  The function may also gain or lose
zeros, but always in pairs.  This can happen either at $\phi_{\rm
top}$ or at a point where $h(\phi_0)$ momentarily develops a double
zero.  Following the motion, appearance, and disappearance of these
zeros provides a framework for the study of the new bounce solutions.

We will start our investigation by considering ``small amplitude''
solutions in which $\phi$ is confined to the region near $\phi_{\rm
top}$, where a linear approximation to Eq.~(\ref{ddotPhiEq}) can be
applied.  We will then follow the evolution of these solutions as a
change of parameters takes them outside the linear regime.  For some
potentials, all bounce solutions arise in this manner.  Other
potentials admit a second type of solution, that are never confined
to the linear region. These can coexist with the former type of
solution, or they may be the only type of bounce present.
 
We begin by introducing an approximation that considerably simplifies
the analysis of the bounce equations.

\subsection{The fixed background approximation}

For the most part, we will focus on
the case where the fractional variation of $V(\phi)$
is small as $\phi$ varies over the range from $\phi_{\rm fv}$ to
$\phi_{\rm tv}$.  This allows us to write
\begin{equation}
   V(\phi) = V_0 + \tilde V(\phi)
\end{equation}
with $|\tilde V(\phi)| \ll V_0$ for all relevant values of $\phi$.
To leading approximation, the curvature of the Euclidean space is then
independent of the value of $\phi$, and we can write 
\begin{equation}
   \rho(\xi) \approx H^{-1} \sin(H\xi) \equiv \rho_0(\xi)
\label{fixedRho}
\end{equation}
with 
\begin{equation}
   H^2 \equiv  {8\pi V_0 \over 3 M_{Pl}^2}  \, .
\end{equation}
If we define $y \equiv H\xi$, the equation for $\phi$ becomes, to
leading approximation, 
\begin{equation}
   {d^2 \phi \over dy^2} + 3\cot y\, {d \phi \over dy} 
        = {1 \over H^2} {d \tilde V \over d\phi}
\label{simplePhiEq}
\end{equation}
with the boundary conditions that $d\phi/dy$ vanish at both $y=0$ and
$y=\pi$.

Once a solution to Eq.~(\ref{simplePhiEq}) has been found, the first
correction to the metric can be obtained by substituting this solution
into a linearized version of Eq.~(\ref{EucFriedmann}) and solving for
$\delta \rho = \rho - \rho_0$.  Because $\rho_0$ is a solution of the
zeroth order problem, the terms linear in $\delta \rho$ do not
contribute to the action.  The terms quadratic in $\delta \rho$ are
suppressed relative to the matter terms in the action by a factor of
order $\tilde V/V_0$, allowing us to write
\begin{equation}
    S_E = - {3 \over 8}\, {M_{Pl}^4 \over V_0 }
       +2\pi^2 \int d\xi \rho_0^3 
        \left({1\over 2} \dot\phi^2 + \tilde V \right) + \cdots \, .
\end{equation}
When the false vacuum action is subtracted from that of the bounce,
the first terms on the right hand sides cancel, leading to 
\begin{equation}
    B =  {9 \over 32} {M_{Pl}^4 \over V_0^2 } \int dy \sin^3 y
       \left[{H^2\over 2} \left({d\phi \over dy}\right)^2 
             +   V(\phi) -  V(\phi_{\rm fv}) \right] \, .
\end{equation}
In particular, for the Hawking-Moss solution
\begin{equation}
    B_{\rm HM} =  {3 \over 8}\, {M_{Pl}^4 \over V_0^2 }
        \left[ V(\phi_{\rm top}) -  V(\phi_{\rm fv}) \right] \, .
\end{equation}

\subsection{Linearized equations}

We begin our analysis by seeking small amplitude solutions where
$\phi_0$ is near $\phi_{\rm top}$ and $\phi$ remains everywhere close
to the top of the barrier.  These are essentially small perturbations
about the Hawking-Moss solution.

We expand the scalar field potential as
\begin{equation}
    \tilde V(\phi) = V(\phi_{\rm top})  + H^2\left[
     - {\beta \over 2} (\phi- \phi_{\rm top})^2 
     +  {b \over 3} (\phi- \phi_{\rm top})^3 
      + {\lambda \over 4} (\phi- \phi_{\rm top})^4
      + \cdots \right] \, .
\label{TaylorV}
\end{equation}
Because $\phi_{\rm top}$ is a maximum of the potential, $\beta$ is
necessarily positive, and
\begin{equation}
    \beta =  {|V''(\phi_{\rm top})| \over H^2} \, .
\end{equation}
The sign of $b$ can be reversed by a simple redefinition of $\phi$, 
and hence is not physically significant.  Finally, $\lambda$ can 
take either sign, although it must be positive if there are no higher
order terms in the potential.

If the field remains sufficiently close to $\phi_{\rm top}$, the cubic
and higher terms in the potential can be dropped in a first
approximation, so that Eq.~(\ref{simplePhiEq}) takes the form
\begin{equation}
  0 =  {d^2 \phi \over dy^2} + 3\cot y\, {d \phi \over dy} 
     + \beta (\phi - \phi_{\rm top}) \, .
\label{linearizedPhi}
\end{equation}
If we write $\phi(y) - \phi_{\rm top} = f(y)\sin^{-3/2}y$,
this becomes
\begin{equation}
   0 = {d^2 f \over dy^2} 
       + \left( \beta + {3\over 2} - {3 \over 4} \cot^2 y \right) f \, .
\end{equation}
Except near the outer edges of the interval $0 \le y \le \pi$, the 
last term in the brackets can be ignored, and we find that 
\begin{equation}
     \phi(y)- \phi_{\rm top} \approx A { \sin(\sqrt{\beta + 3/2} \, y
               + \delta) \over \sin^{3/2}y } \, ,
\end{equation}
with $A$ and $\delta$ constants.

More precisely, we can note that an exact solution of
Eq.~(\ref{linearizedPhi}) is given by a Gegenbauer, or ultraspherical,
function
\begin{equation}
    \phi(y)- \phi_{\rm top} = A C_\alpha^{3/2}(\cos y)
\end{equation}
with $\alpha$ the positive root of 
\begin{equation}
     \alpha(\alpha +3) = \beta \, .
\end{equation}
The fact that $C_\alpha^{3/2}(1)$ is finite for all real positive
$\alpha$ guarantees the vanishing of $d\phi/dy$ at $y=0$; it is this
condition that eliminates the Gegenbauer function of the second kind,
$D_\alpha^{3/2}(\cos y)$.  The second boundary condition, that
$d\phi/dy$ also vanish at $y=\pi$, is only satisfied if $\alpha$ is
an integer.  Hence, the linearized approximation,
Eq.~(\ref{linearizedPhi}), has an acceptable solution only if
$V''(\phi_{\rm top}) = N(N+3) H^2$ for $N=1,2,\dots$.  Because
$C_N^{3/2}(\cos y)$ has $N$ zeros in the range $0 \le y \le \pi$,
these solutions are bounces with with $k=N$ crossings of the barrier.

We recall, for later use, that the Gegenbauer functions for integer
$N$ are polynomials.  If defined with the standard normalization
\begin{equation}
    C_N^{3/2}(1) = {(N+1)(N+2) \over 2} \, ,
\end{equation}
they obey the orthogonality relation
\begin{equation}
   \int_{-1}^1 \, dx \, (1-x^2) C_M^{3/2}(x) C_N^{3/2}(x)
      = \delta_{M,N} \, {2(N+1)(N+2) \over (2N+3) }  
          \equiv \delta_{M,N} \,  k_N \, .
\label{gegennorm}
\end{equation}

\subsection{Incorporating nonlinearities}
\label{firstnonlinear}

The linearized equations only have solutions obeying the boundary
conditions when $\beta$ is equal to one of a discrete set of critical
values.  This condition is relaxed once the nonlinear terms are
included.  This is quite similar to the way in which adding anharmonic
terms to a simple harmonic oscillator allows small amplitude
oscillations with a continuous range of periods.  Just as the
amplitude is related to the period in the oscillator problem, the
amplitude of the bounce solutions is determined by the distance
$\Delta = \beta - N(N+3)$ from the critical value.

Because the Gegenbauer polynomials form a complete set, an
arbitrary function $\phi(y)$ obeying the boundary conditions of
Eq.~(\ref{curvedBdy}) can be expanded as
\begin{equation}
    \phi(y) = \phi_{\rm top} + {1 \over \sqrt{|\lambda|}} 
                \sum_{M=0}^\infty A_M C_M^{3/2}(y)
		\label{gegenExpansionOfPhi} \, ,
\end{equation}
where the coefficients $A_M$ are dimensionless.
Substituting this into Eq.~(\ref{simplePhiEq}) and retaining the
contributions from the cubic and quartic terms in the expansion of the
potential yields
\begin{equation}
   0 = \sum_{M=0}^\infty C_M^{3/2}(y) \left\{ [\beta - M(M+3)] A_M 
      - {b \over \sqrt{|\lambda|}}  \sum_{I,J} A_I A_J p_{IJ;M} 
      - {\rm sgn}\,(\lambda) \sum_{I,J,K} A_I A_J A_K q_{IJK;M}
   \right\} 
\label{gegenExpansion}
\end{equation}
where the terms involving
\begin{equation}
   p_{IJ;M} = k_M^{-1/2} \int_{-1}^1 dy \, C_I^{3/2}(y) C_J^{3/2}(y)
              C_M^{3/2}(y)
\end{equation}
and
\begin{equation}
   q_{IJK;M}  = k_M^{-1/2}\int_{-1}^1 dy \, C_I^{3/2}(y) C_J^{3/2}(y)
              C_K^{3/2}(y) C_M^{3/2}(y) 
\end{equation}
arise from the expansion of the $\phi^2$ and $\phi^3$ terms, and the
normalization factor $k_M$ is defined by Eq.~(\ref{gegennorm}).  Note
that $p_{IJ;M}$ vanishes if the sum of any two indices is greater than
the third, and that $q_{NNN;N}>0$ for all $N$.

Each term of the sum in Eq.~(\ref{gegenExpansion}) must vanish
separately.  For $|\Delta|$ sufficiently small, we expect to find a
small amplitude solution in which a single coefficient, $A_N$, is
dominant.  Explicitly,
\begin{equation}
    0 = \Delta A_N - {b \over \sqrt{|\lambda|}}  p_{NN;N} A_N^2 
       - {2b \over \sqrt{|\lambda|}} \sum_{M\ne N} p_{MN;N} A_M A_N
          -  {\rm sgn}\,(\lambda) q_{NNN;N} A_N^3 +\cdots
\end{equation}
and 
\begin{equation}
    0 = [\beta - M(M+3)] A_M -  {b \over \sqrt{|\lambda|}} \, p_{NN;M} A_N^2 
          + \cdots \, , \qquad M\ne N   \, ,
\end{equation}
where in each equation the omitted terms are higher order in $A_N$.
Using the second equation to eliminate $A_M$ from the first leads to
\begin{equation}
   0 = \Delta - {b \over \sqrt{|\lambda|}}  p_{NN;N} A_N
                     - c A_N^2 + O(A_N^3) \, ,
\end{equation} 
where 
\begin{equation}
    c = {\rm sgn}\,(\lambda) q_{NNN;N} 
      + {2b^2\over |\lambda|}
      \sum_{M\ne N}^{2N}\left[{1 \over \beta - M(M+3)}\right] p_{NN;M}
      p_{NM;N}  \, .
\end{equation}
This gives
\begin{equation}
   A_N = {1 \over 2c} \left[ - {b p_{NN;N} \over \sqrt{|\lambda|}}
      \pm \sqrt{ {b^2(p_{NN;N})^2 \over |\lambda|} + 4 c \Delta }
      \right] \, .
\label{AnSolutions}
\end{equation}

If the cubic term in Eq.~(\ref{TaylorV}) vanishes, $b=0$, there are
real solutions for $A_N$ only if $c \Delta >0$, with $A_N = \pm
\sqrt{\Delta/c}$.  Thus, for $\lambda >0$ there are no small amplitude
bounces if $\beta < N(N+3)$.  As $\beta$ is increased past this
critical value, two solutions appear.  For odd $N$ these are a
physically equivalent true-to-false and false-to-true pair, while for
even $N$ they are two distinct solutions, one true-to-true and the
other false-to-false.  Things are similar if $\lambda <0$, except that
the solutions only exist for $\beta < N(N+3)$.

For nonzero $b$ and odd $N$, the situation is almost the same.  Because
$p_{NN;N}$ vanishes for odd $N$, the condition for existence of a
solution is again that $c \Delta > 0$, although now $c$ and $\lambda$
may have different signs.  As before, there are two distinct but
physically equivalent small amplitude solutions, and these appear at
the critical value $\beta = N(N+3)$.

For nonzero $b$ and even $N$, the situation is more complicated.
From Eq.~(\ref{AnSolutions}), we see that the critical value of $\beta$
where solutions first appear is not $\beta = N(N+3)$, but rather
\begin{equation}
    \beta = N(N+3) - {b^2 \over 4c|\lambda|} (p_{NN;N})^2 \, ,
\label{betacrit}
\end{equation}
which is a bit smaller (larger) if $c$ is positive (negative).  In
contrast with the previous cases, 
when the solutions first appear they have a finite amplitude and
both have the same sign for $A_N$.  As $\beta$ is increased
(decreased) through $N(N+3)$, one of the solutions passes through zero
and then takes on the opposite sign.  After this point the two
solutions remain of opposite sign, with the magnitude of $A_N$
increasing for both.

Some words of caution are in order here.  Our analysis is valid only
for sufficiently small $A_N$.  If $b/\sqrt{|\lambda|}$ is too large,
this condition may not be met when $\beta$ is given by
Eq.~(\ref{betacrit}), in which case it will never be satisfied by the
larger of the two solutions.  However, as $\Delta \rightarrow 0$, the
other solution for $A_N$ tends to zero, and so should be reliable.  As
discussed at the beginning of this section, the zeros
of $h(\phi_0)$ appear or disappear in pairs.  Hence, the
existence of one small amplitude solution implies the existence of the
second.  We therefore expect the qualitative features --- the
emergence of two solutions, with finite amplitude, at a value of $\beta$
not corresponding to an integer --- to be valid even if the
quantitative analysis is not completely reliable.

The results of this section can be compared with the analysis of
Jensen and Steinhardt\cite{Jensen:1983ac}.  They argued that if $|V''|$ is
monotonically decreasing as one moves from the top of the potential
toward the false vacuum there will be a unique $k=1$ Coleman-De Luccia
bounce if (in our language) $\beta > 4$, and that this solution merges
with the Hawking-Moss solution as $\beta \rightarrow 4$.  When applied
to Eq.~(\ref{TaylorV}), their condition on $V''$ translates into the
statement that $\lambda > 0$.  If $b^2/\lambda$ is not too large, $c$
will also be positive, and our results will agree with theirs.
However, for $b^2/\lambda$ sufficiently large, $c$ can be negative even
though $\lambda$ is not.  In this case, $\beta = 4$ is an upper bound
for the existence of these solutions, and our results will disagree
with their analysis.  We believe that this can be attributed to the
fact that their claims are based on the argument that the potential
gets ``flatter'' as one moves away from the top of the barrier; for
large $b$ this is true, in the relevant range, only on one side of the
potential.

Perturbative methods similar to those used in this section have
recently been applied to the calculation of the bounce
action~\cite{Balek:2004sd}.

\section{Numerical results for $\lambda > 0$}

Once the amplitude become large, the analysis based on a single
dominant term in Eq.~(\ref{gegenExpansionOfPhi}) fails, and we must
resort to numerical methods.  Before describing our numerical results,
we outline some qualitative arguments that may provide some useful
insight.  We focus here on the case $\lambda > 0$, where the quartic
term in the potential is positive.  We will return to the $\lambda <
0$ case in Sec.~\ref{flatbounce}.

\begin{figure}[t]
\begin{center}
\epsfig{file=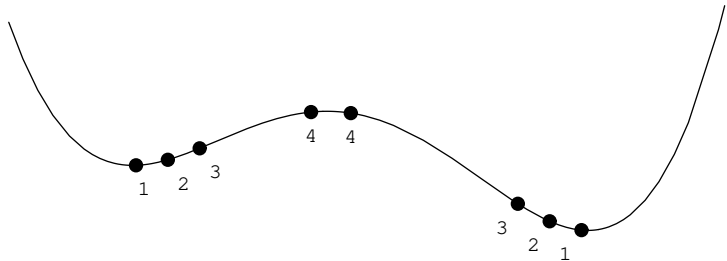, width =10cm }
\end{center}
\caption{A schematic plot of the starting points of bounce solutions
to a $\lambda > 0$ theory, such as that discussed in Sec.~V, with
$\beta$ slightly greater than 28.  The numbers next to the points
represent the number of times the solution crosses the top of the
potential barrier.
\label{stdZeros}}
\end{figure}

We have seen that $h(\phi_0)$ has a zero at $\phi_{\rm top}$,
corresponding to the Hawking-Moss solution, for all values of $\beta$.
As $\beta$ increases, pairs of new zeros, corresponding to bounce
solutions with ever increasing values of $k$, appear at $\phi_{\rm
top}$ as $\beta$ passes through successive critical values.  When the
analysis of Sec.~\ref{firstnonlinear} can be applied, these zeros move
apart as $\beta$ increases.  We expect this behavior to continue even
after the zeros have moved beyond the small amplitude regime.  Thus,
at any given $\beta$ there should be a sequence of bounce solutions.
For $b=0$ and $N(N+3) < \beta < (N+1)(N+4)$, there would be two
solutions each for $k = 1, 2, \dots, N$, with the two solutions for
each odd $k$ simply being ``$y$-reversed'' versions of each other.
For nonzero $b$ the picture would be similar, except that the new
solutions for even $N$ appear somewhat before $\beta = N(N+3)$.
Figure~\ref{stdZeros} gives a schematic illustration of the expected
pattern of zeros.

We can also make some predictions about the form of these solutions.
It seems reasonable to expect that, even after $\phi_0$ has moved well
beyond the small amplitude regime, the average period of the
oscillations will continue to be determined to leading order by
$\beta$.  Increasing $\beta$ will decrease this period, and so should
cause a plot of the oscillations of $\phi$ to shrink horizontally at
the same time that it is expanding vertically.  With the number of
oscillations in a given solution remaining fixed, this will open up
``gaps'' at the ends of the interval.  In order to satisfy the
boundary conditions that $d\phi/dy =0$ at $y=0$ and $y=\pi$, the field
must be relatively constant in these gaps. This, in turn, implies that
the field near the end points must be close to one or the other of the
minima of the potential.  As $\beta$ is increased further, the
fraction of the $y$-interval occupied by oscillations will continue to
decrease, while the regions occupied by these false and true vacuum
regions will expand.

We have tested these ideas numerically by considering a theory governed
by a quartic potential than can be written in the form
\begin{equation}
    V = \beta H^2 v^2 \left[-{1\over 2} \psi^2 - {g \over 3}\psi^3 
       + {1\over 4}\psi^4 \right]  + V_0  \, ,
\end{equation}
where $\psi = \phi/v$ is a dimensionless field that has been rescaled
so that for $g=0$ the minima of the potential are at $\psi = \pm 1$;
in the notation of Eq.~(\ref{TaylorV}), this corresponds to
$b=-g\beta/2v$ and $\lambda = \beta/v^2$.  We focus here on
our results for a single value of $\beta$.  However, we have explored
solutions for a much wider range of values, going as high as $10^4$.
In all cases, our results are consistent with the picture
we describe here.

\begin{figure}[thp]
\begin{center}
\epsfig{file=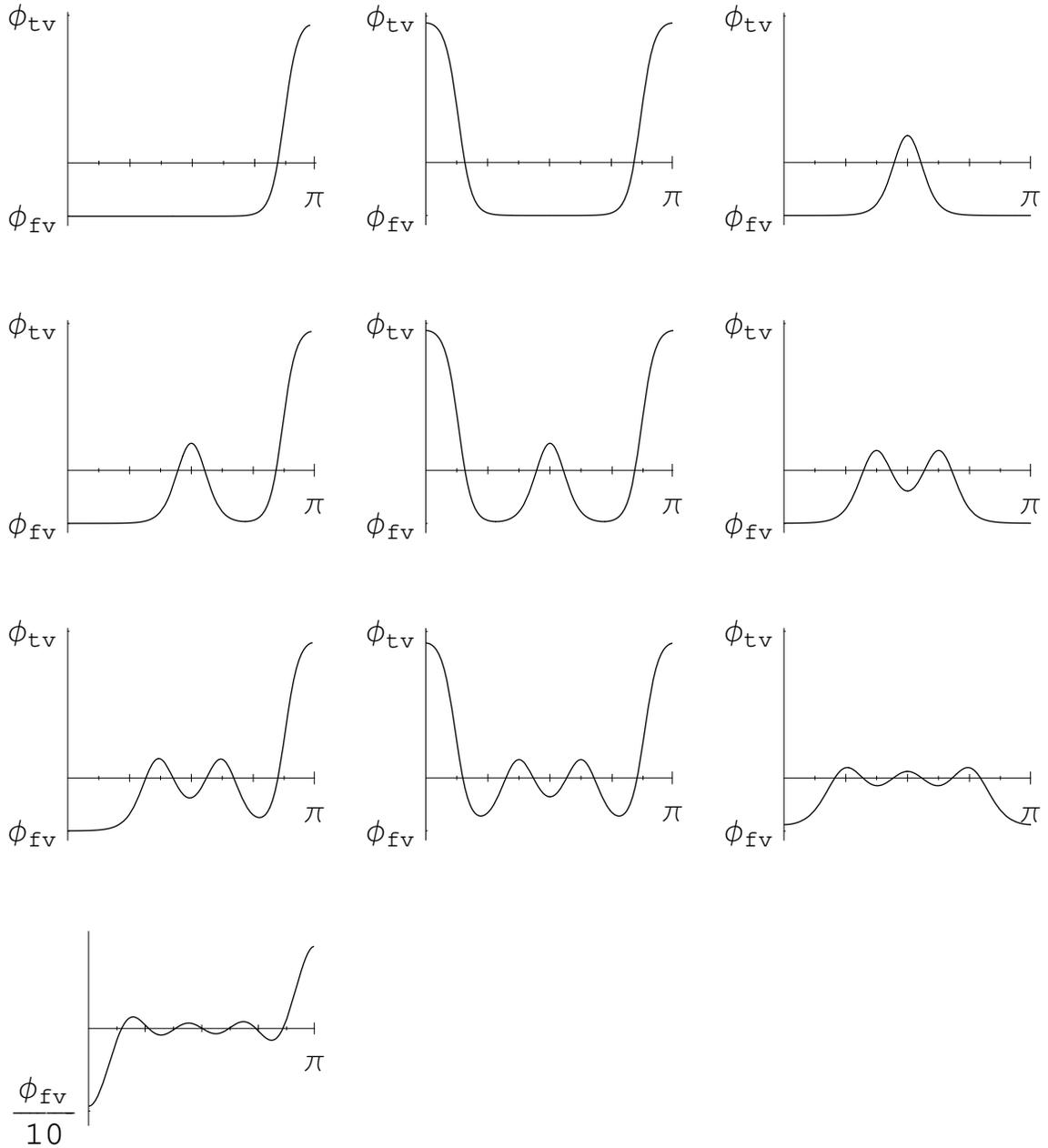, width =16cm }
\end{center}
\caption{Bounce solutions for the asymmetric ($g = {1
\over 2 \sqrt{2}}$) potential discussed in the text.  For these
solutions $\beta = 70.03$.
   \label{asymArray}}
\end{figure}

In Fig.~\ref{asymArray} we show the bounce solutions for
$g =1/2\sqrt{2}$ and $\beta = 70.03$, corresponding to $N=7$.  As
expected, we find solutions for every integral value of $k$ up to
$k=7$, with two distinct solutions for each even value of
$k$.  Although it is barely apparent on the figures (except for
$k=7$), it should be noted that, among solutions starting on a given
side of the barrier, the initial value $\phi_0$ monotonically
approaches $\phi_{\rm top}$ as $k$ increases, in accord with our
expectations.

The $k=1$ solution is the Coleman-De Luccia bounce.  For $k=2$, there
are two solutions.  One starts and ends near the true vacuum, and
might be viewed as a two-bounce solution analogous to the flat space
multibounce solutions that are encountered in the path integral
treatment.  Indeed, its field profile is very close to that expected
from two independent bounces and its action is approximately twice
that of the single bounce.  The other $k=2$ solution starts and ends
near the false vacuum.  It, too, might be interpreted as a two-bounce
solution, if one took the Coleman-De Luccia bounce to
represent decay from the true vacuum to the false.  From this
viewpoint, the false vacuum region corresponds to the bubble interior;
since this is large compared to the size of the four-sphere, there
must be significant distortion to be able to fit in two bounces.  Note
that even in the region separating the two ``interiors'' $\phi$ does
not get very close to its true vacuum value.

The solutions with $k>2$ all have intermediate oscillations about
$\phi_{\rm top}$.  As expected, the frequency of these oscillations
seems to be determined primarily by $\beta$, with the separation
between zeros of the field being roughly independent of $k$. 
The existence of intermediate oscillations seems to have little
effect on the true vacuum bubble region, with the field profiles near
$y=\pi$ for the $k<7$ false-to-true solutions and near $y=0$ and
$y=\pi$ for all the true-to-true solutions being very similar to that
for the corresponding region of the Coleman-De Luccia solution.  This
is not so for the false vacuum regions.

The plot for the $k=7$ solution shows that the linear approximation
works quite well, as should be expected with $\beta$ so close to the
critical value of 70.  The numerical solution is indistinguishable to
the eye from the Gegenbauer polynomial, and the initial value,
$\phi_0$, agrees well with the prediction of Eq.~(\ref{AnSolutions}).

\begin{figure}[thp]
\begin{center}
\epsfig{file=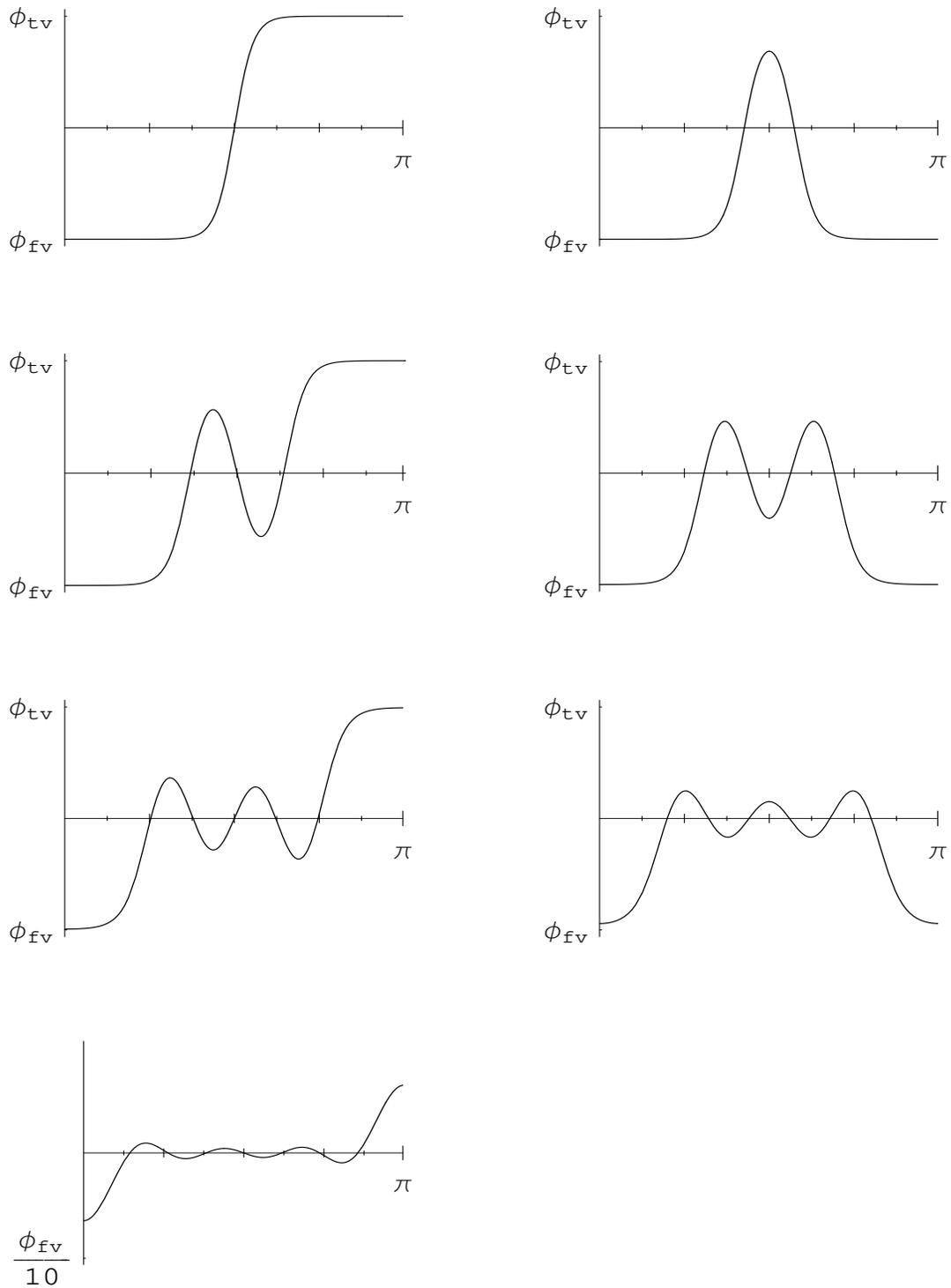, width =15cm, height =21cm}
\end{center}
\caption{Bounce solutions for the symmetric ($g=0$)
potential with two degenerate vacua that is discussed in the text.
Again, $\beta = 70.03$.  
  \label{symArray}}
\end{figure}

In Fig.~\ref{symArray}, we show the analogous results for the case of
$g=0$, where the two minima are degenerate.  In flat spacetime,
there would be no bounces for this case but, as we see, this is no
longer so when gravitational effects are included.

\begin{figure}[htp]
\begin{center}
\epsfig{file=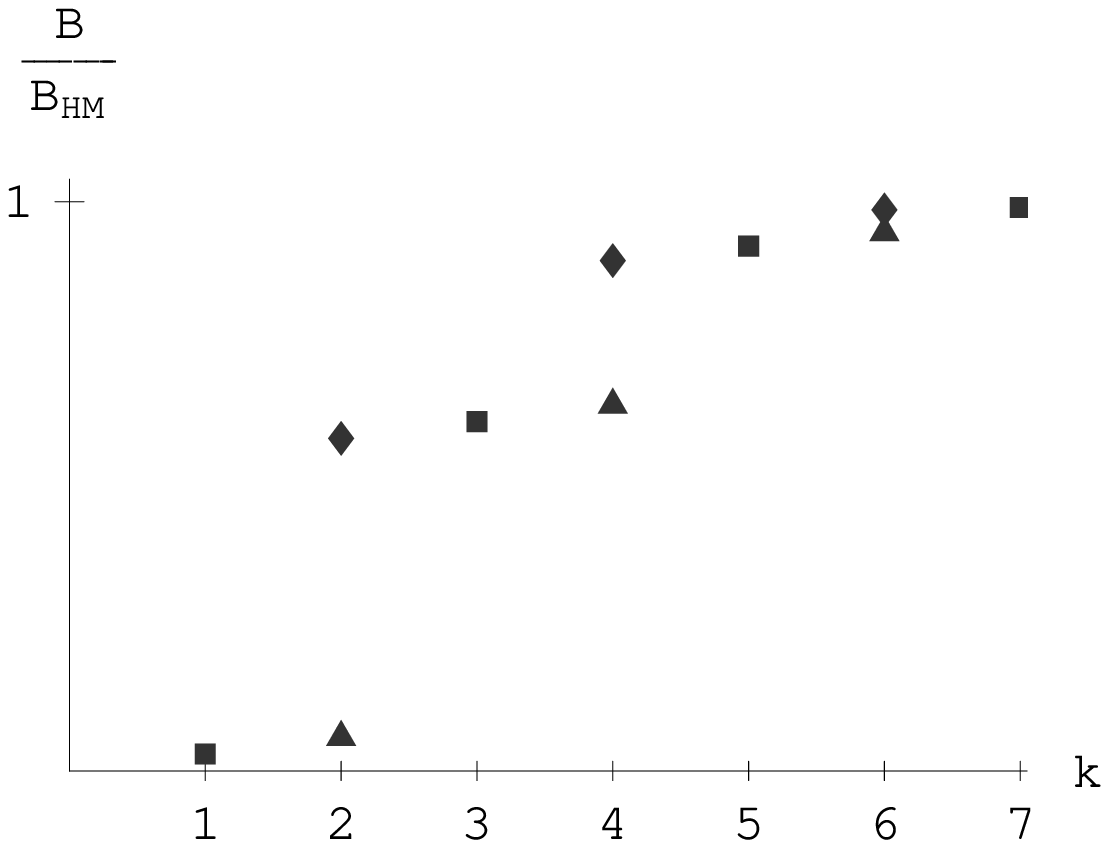, width =10cm}
\end{center}
\caption{The tunneling exponent $B=S(\phi_{\rm bounce}) - S(\phi_{\rm
    fv})$ for the asymmetric potential bounce solutions shown in
    Fig.~4.  The values are given as fractions of the Hawking-Moss
    value, and are plotted as a function of the number of times the
    solution crosses the top of the barrier.  Boxes correspond to
    solutions which interpolate between the true and false vacuum
    sides of the barrier, diamonds correspond to ``true-to-true''
    solutions, and triangles to ``false-to-false'' ones.
\label{asymAction}}
\end{figure}

\begin{figure}[htp]
\begin{center}
\epsfig{file=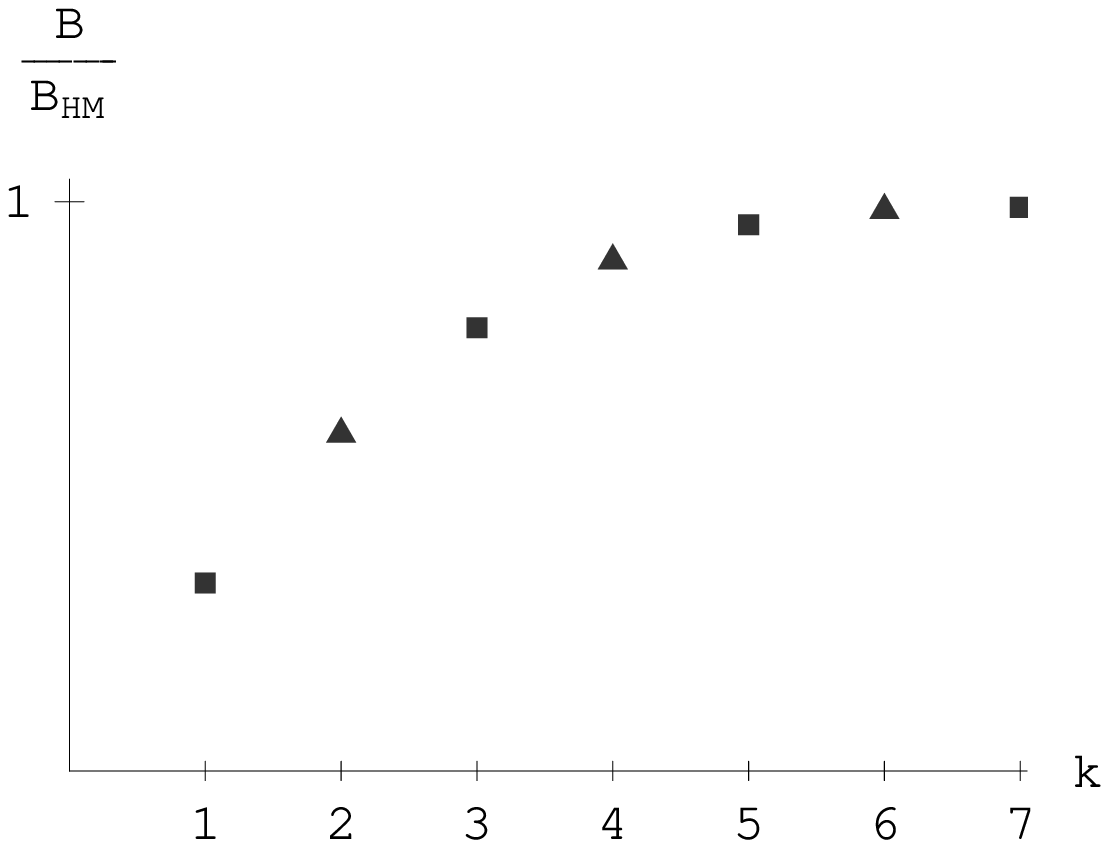, width =10cm}
\end{center}
\caption{The tunneling exponent $B$ for the symmetric potential bounce
   solutions shown in Fig.~5.  The values are given as fractions of
   the Hawking-Moss value, and are plotted as a function of the number
   of times the solution crosses the top of the barrier. Boxes
   correspond to solutions which interpolate between the the two
   vacua, and triangles correspond to solutions which begin and end on
   the same side of the barrier. 
\label{symAction}}
\end{figure}

Finally, in Figs.~\ref{asymAction} and \ref{symAction} we show the
actions for these solutions.  (More precisely, what we actually plot
in these figures is $B= S_{\rm bounce} - S_{\rm fv}$, which is the
quantity that would appear in the exponent if these solutions were
interpreted as contributions to the process of tunneling out of the
false vacuum.)  As $k$ increases, $B$ approaches the Hawking-Moss
value $B_{\rm HM}$.  This can be understood by noting that in the
linearized approximation the contribution from the gradient terms
exactly cancels that from the oscillatory part of the potential term,
so that the total action is exactly equal to that of the Hawking-Moss
solution.

\section{Bounces with flat potential barriers}
\label{flatbounce}

Our analysis of small amplitude bounces in the previous two sections
showed that for a wide class of potentials --- those with a positive
fourth derivative, and not too large a third derivative, at $\phi_{\rm
top}$ --- there is neither a small oscillation bounce nor a bounce
smoothly related to a small oscillation bounce if $|V''(\phi_{\rm
top})| < 4H^2$.  This result is intuitively quite plausible.  One
might expect $|V''(\phi_{\rm top})|^{-1/2}$ to set a natural scale for
the size of the bounce, so that if this were too large, relative to
the radius $H^{-1}$, the bounce would not fit on the four-sphere.

However, our results for the case where the fourth derivative (or more
precisely, the related quantity $c$) is negative show that this
seemingly plausible argument cannot be quite correct.  For this case we 
saw that there was a small amplitude $k=1$ bounce
only for a range of values extending {\it downward} from $\beta=4$.
One logical possibility is that these solutions continue (although
not necessarily with small amplitudes) all the way down to $\beta=0$.
Alternatively, there might be some minimum value $\beta_{\rm min}$
(which could depend on the other parameters of the theory) for which
the bounce exists. Since solutions can only appear and disappear in
pairs, this latter possibility would require the existence of a second
solution, which would also appear at $\beta_{\rm min}$ but which would
persist beyond $\beta = 4$.

In this section we will explore this direction in more detail.  Because
the potentials that evade the $\beta >4$ bound are characterized by
being relatively flat at the top of the barrier, we begin by studying
a toy model, defined by the potential
\begin{equation}
   V = \cases  {C (\phi + a) + V_0  \, ,\qquad \phi < -a \cr \cr
             V_0  \, ,\qquad -a \le \phi \le a \cr \cr
         -C (\phi - a) + V_0  \, ,\qquad \phi > a }
\label{toypotential}
\end{equation}
that is absolutely flat at the top.  This potential has the additional
advantage that the field equation can be solved
analytically.\footnote{The fact that this potential has no minima is
irrelevant for our purposes, since the bounce solutions never quite
reach either the true or false vacuum in any case.  One could, of
course, modify the potential to give minima at large values of
$|\phi|$.}

Inserting this potential into Eq.~(\ref{simplePhiEq}), and defining
\begin{equation}
    \chi(y) = \sin^3 y {d\phi\over dy} \, ,
\end{equation}
leads to
\begin{equation}
    {d\chi\over dy}
       = \cases { a \gamma \sin^3 y   \, ,\qquad \phi < -a \cr\cr
              0  \, ,\qquad -a \le \phi \le a \cr\cr
            -a \gamma \sin^3 y \, ,\qquad \phi > a }
\label{chiEq}
\end{equation}
where 
\begin{equation}
      \gamma = {C \over a H^2} \, .
\end{equation}
Because $C/a$ gives a rough measure of the average value of $|V''|$
for our toy potential, $\gamma$ is somewhat
analogous to the quantity $\beta$ that appeared in the analysis of the
two previous sections.

We first seek a $k=1$ (Coleman-De Luccia) bounce.  The symmetry of the
potential implies that this solution (as well as any other bounce with
odd $k$) must have $\phi(\pi/2) =0$.  Let $\phi(0) = \phi_0 < -a$, and
define $y_1 < \pi/2$ by the condition $\phi(y_1) = -a$.
Integrating Eq.~(\ref{chiEq}) over the interval $0 \le y \le y_1$ gives
\begin{equation}
   \chi(y_1) = a \gamma \int_0^{y_1} du \sin^3 u \, .
\end{equation}
The fact that $\chi$ is constant over the flat part of the potential
implies that
\begin{equation}
   a = \phi(\pi/2) - \phi(y_1) 
            = \chi(y_1)  \int_{y_1}^{\pi/2} {dv \over \sin^3 v} \, .
\label{k1flatPart}
\end{equation}
Combining these two equations gives the requirement that 
\begin{equation}
       {1 \over \gamma} = \int_0^{y_1} du \sin^3 u 
        \int_{y_1}^{\pi/2} {dv \over \sin^3 v} \equiv F_1(y_1) \, .
\label{k1toyCondition}
\end{equation}
Once a value of $y_1$ satisfying this condition has been found, it is
a straightforward matter to find $\phi_0$ by integrating back from
$y_1$ to $y=0$ and to then integrate forward to obtain $\phi(y)$ for
all $y$.

\begin{figure}[tbp]
\begin{center}
\epsfig{file=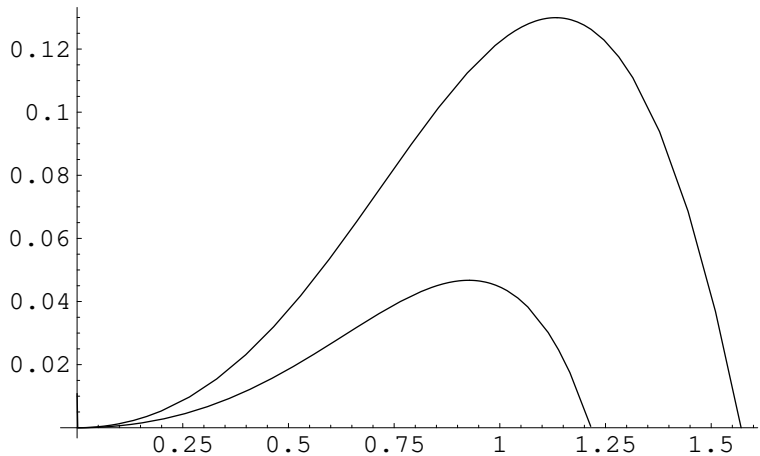, width =10cm}
\end{center}
\caption{The functions $F_1(y_1)$ and $F_2(y_1)$ for the toy model defined
  by Eq.~(\ref{toypotential})
  \label{F1plot}} 
\end{figure}

\begin{figure}[tbp]
\begin{center} 
\epsfig{file=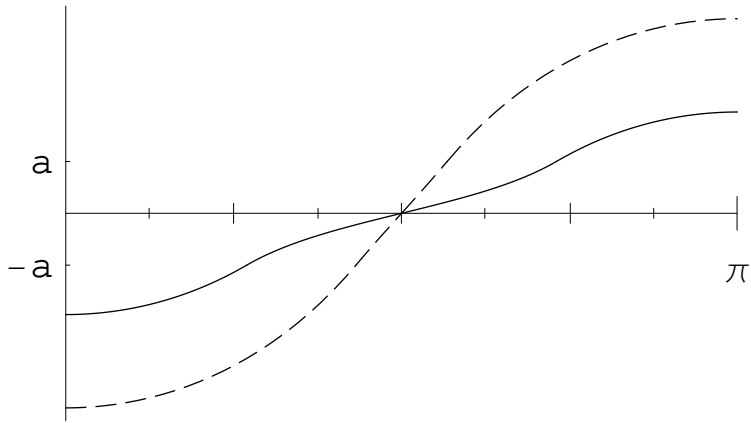, width =10cm}
\end{center}
\caption{The two $k=1$ bounces for our linear toy model when
 $\gamma=10$.}
\label{toybounces}
\end{figure}

From the plot of $F_1(y)$ shown in Fig.~\ref{F1plot}, we see that
Eq.~(\ref{k1toyCondition}) has no solutions, and hence there is no
Coleman-De Luccia bounce, if $\gamma < 1/{\rm Max}\, [F_1(y)] \approx
7.70$.  For any $\gamma$ greater than this critical value there are
two acceptable values for $y_1$, and thus two distinct Coleman-De
Luccia bounces.  As examples of this, in Fig.~\ref{toybounces} we
show the two bounces solutions for $\gamma=10$.

In general, these values for $y_1$ can only be found
numerically.  However, approximate expressions can be obtained for the
limiting case $\gamma \gg 1$.  In one solution, with
\begin{equation}
      y_1 \approx \sqrt{8 \over \gamma} \qquad\qquad \phi_0 \approx 
                 -a \left( 1 + {1 \over \gamma}\right) \, ,
\end{equation}
the field is mostly on the flat part of the
potential.  In the other solution, with
\begin{equation}
     y_1 \approx {\pi \over 2} - {3 \over 2\gamma}  \qquad\qquad 
        \phi_0 \approx -\left({1 + \ln 4 \over 6}\right) a\gamma \approx
        -.398 a\gamma \, ,
\end{equation}
the field is mostly on the sloping parts of the potential.  The 
latter solution has the lower action.

For $k=2$ (or any even value of $k$), $d\phi/dy$ vanishes at
$y=\pi/2$.  Let us define $y_1 < y_2 < \pi/2$ by 
$\phi(y_1) = - \phi(y_2) = -a$.  Integrating Eq.~(\ref{chiEq})
then gives
\begin{equation}
   \chi(y_1) = a\gamma \int_0^{y_1} du \sin^3 u
\end{equation}
\begin{equation}
   \chi(y_2) =  a\gamma\int_{y_2}^{\pi/2} du \sin^3 u \, .
\end{equation}
Because the field is in the flat part of the potential for $y_1 < y <
y_2$, these integrals must be equal; this implicitly defines
$y_2$ as a function of $y_1$.  

By analogy with Eq.~(\ref{k1flatPart}), we also have 
\begin{equation}
   2a = \phi(y_2) - \phi(y_1) 
            = \chi(y_1) \int_{y_1}^{y_2(y_1)} {du \over \sin^3 u}
      \equiv 2 a\gamma F_2(y_1) \, .
\end{equation}
Referring again to Fig.~\ref{F1plot}, we see that there are no
solutions for $\gamma <1/ {\rm Max}\, [F_2(y)] \approx 21.7$, but two
solutions for all larger values of $\gamma$.  In the cases where there
are solutions, these have values of $y_1$ (and also of $\phi_0$) that
fall between the corresponding $k=1$ values.  In the limit $\gamma \gg 
1$, the $k=2$ solutions have
\begin{equation}
      y_1 \approx  \sqrt{16 \over \gamma}
         \qquad\qquad 
      \phi_0 \approx -a \left( 1 + {2 \over \gamma}\right)
\end{equation}
and 
\begin{equation}
      y_1 \approx 1.22
          \qquad\qquad 
      \phi_0 \approx -.319 a\gamma \, .
\end{equation}

This procedure can be extended to all higher values of $k$.  For any
given $k$, solutions only exist if $\gamma$ is larger than a critical
value $\gamma_{\rm cr}(k)$; if $\gamma > \gamma_{\rm cr}(k)$ there are
two allowed values for $\phi_0$, both of which fall between the
allowed valued for the $k-1$ solutions with the same $\gamma$.  For
$k=3$, the critical value is 42.7.  More generally, $\gamma_{\rm
cr}(k)$ is an increasing function of $k$ and is $O(k^2)$ for large
$k$; this behavior is reminiscent of the critical values of $\beta$
that govern the appearance of small amplitude solutions in potentials
with a positive quartic term.  

\begin{figure}[tbp]
\begin{center}
\epsfig{file=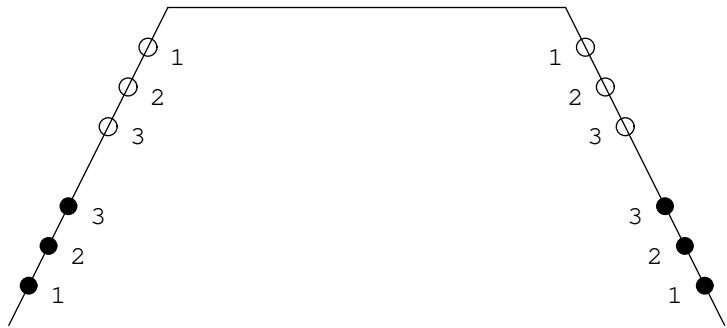, width =10cm}
\end{center}
\caption{A schematic graph of the starting points of bounce solutions
to our linear toy model.  Closed circles represent solutions which
move outward as $H$ is decreased, and open circles represent solutions
which move inward.  The numbers next to the points represent the
number of times the solution crosses the top of the potential barrier.
\label{toyzeros}}
\end{figure}

The situation is illustrated in Fig.~\ref{toyzeros}, which gives a
schematic summary of the possible values for $\phi_0$ for a given
value of $\gamma$.  As $\gamma$ is increased, the zeros marked by
closed circles move outward, while those marked by open circles move
inward.  When $\gamma$ reaches a critical value, two new pairs of
coincident zeros (one pair on each side of the potential) appear and
then begin to separate.  With increasing $\gamma$, accumulations of
open circles develop near $\phi=\pm a$, while the outwardly moving
closed circles become further and further apart.

The actions of these solutions may be either greater or less than that
of the Hawking-Moss solution.  For example, the $k=1$ solution with the
smaller $|\phi_0|$ always has a higher action than both the other
$k=1$ solution and the Hawking-Moss.  The action of the other $k=1$
solution is greater than $S_{\rm HM}$ for small $\gamma$, but not for
large $\gamma$.

This toy model has the advantages of allowing analytic solution and of
emphasizing that it is possible to have a Coleman-De Luccia bounce
with an arbitrarily small value of $|V''(\phi_{\rm top})|$.  However,
one consequence of its somewhat artificial form is that the bounce
solutions have some special properties that we would not expect in a
more generic potential.  For example, the values of $\phi_0$ that move
inward with increasing $\gamma$ (i.e., those indicated by open circles
in Fig.~\ref{toyzeros}) accumulate near the ``corners'' of the
potential at $\pm a$, but never move further inward where they might
meet and ``annihilate''.

A less artificial example can be obtained by considering a potential
of the form $V = -a\phi^2 - b\phi^4 +c\phi^6$, with $a$, $b$, and $c$
all positive.  With three parameters to vary, it is possible to make
the potential relatively flat, as defined by a scale-independent
criterion, at the top of the barrier.  For example, if $\pm v$ are the
minima of the potential, we can define an averaged second derivative
$V''_{\rm avg}$ by
\begin{equation}
    V''_{\rm avg} =  {V'(v/2)  - V'(-v/2) \over v}  \, .
\end{equation} 
By choosing $b^2 \gg ac$ we can make $|V''(0)| \ll |V''_{\rm avg}|$.

We have followed the evolution of the bounce solutions for such a
potential (with $b^2/ac= 80$) as $H$ is varied with the parameters in
the potential held fixed.  For very large $H$, there are no bounce
solutions, in analogy with the low $\beta$ or low $\gamma$ behavior of
our previous examples.  As $H$ is decreased, two pairs of zeros,
corresponding to initial values for $k=1$ bounces, appear at nonzero
points on opposite sides of the barrier, just as in our toy model.
(In the example we studied, these first appeared at $|V''_{\rm avg}| /
H^2 =4.03$.)  As $H$ is reduced further, each pair splits, with one
zero moving inward and one outward.\footnote{Another example whose
$k=1$ solutions display a similar behavior was discussed by Jensen and
Steinhardt~\cite{Jensen:1988zx}.}
With further decreases in $H$,
additional zeros, corresponding to bounces with $k=2$, $k=3$,
etc. appear, just as in the toy model.

\begin{figure}[t]
\begin{center}
\epsfig{file=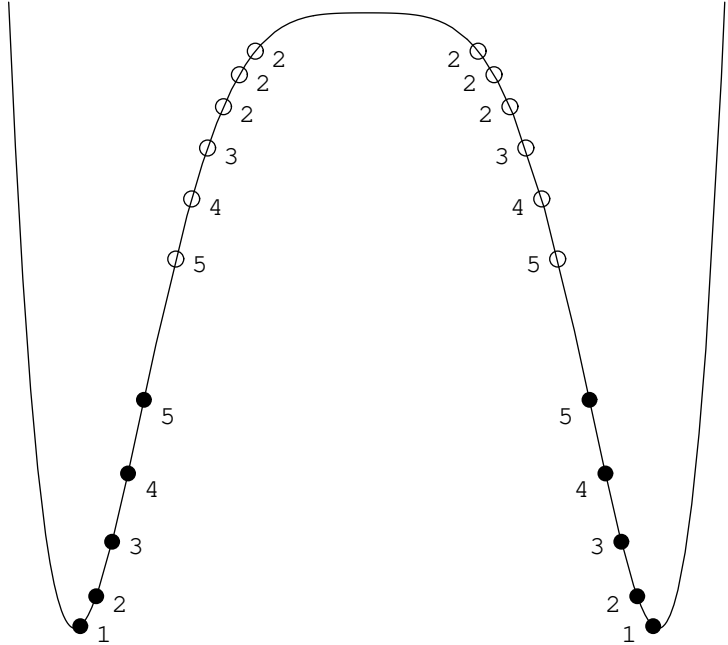, width =10cm}
\end{center}
\caption{A schematic graph of the starting points of bounce solutions
for the sixth-order potential discussed in the text.  Closed circles
represent solutions which move towards one of the vacua as $H$ is
decreased, and open circles represent solutions which move towards the
top.  The numbers next to the points represent the number of times the
solution crosses the top of the potential barrier.
\label{phisixzeros}}
\end{figure}

In contrast with the toy model, the inwardly moving zeros proceed all
the way in to $\phi=0$, reaching that point exactly when predicted by
the analysis of Sec.~\ref{firstnonlinear}.  Thus, two $k=1$ zeros meet
and annihilate when $H^2 = a/2$, two $k=2$ zeros annihilate when $H^2
= a/5$, etc.  We have also encountered a more complex type of behavior
for some values of the parameters.  As described above, two pairs of
$k=2$ zeros appear at a certain critical value of $H$.  After moving
inward some distance, each of the inner zeros splits into three zeros.
These then separate for a time, and then rejoin to form a single zero,
which eventually moves to $\phi=0$.  A schematic snapshot of this
example for a fixed value of $H$ is shown in Fig.~\ref{phisixzeros}.

The examples considered in this section clearly demonstrate that the
existence of a Coleman-De Luccia bounce need not impose a lower bound
on $|V''(\phi_{\rm top})|$.  However, they suggest that a more
generalized version of this bound, involving an averaged value of
$|V''|/H^2$, does hold.  This is consistent with the results
of~\cite{Balek:2003uu}, where it is argued that a necessary condition
for the existence of a Coleman-De Luccia bounce is that
$|V''(\phi)|/H^2 > 4$ for some value of $\phi$.

\section{Concluding remarks}
\label{conclusion}

In this concluding section we will first summarize our results and
then discuss how they might be affected if we relax some of the
restrictions that we have imposed to simplify the analysis.  We will
then briefly discuss the physical implications and interpretation of
these solutions.

\subsection{Summary}

In this paper we have studied in some detail a class of oscillating
bounce solutions that arise when the effects of gravity on vacuum
tunneling processes are taken into account.  These are qualitatively
distinct from the Coleman-De Luccia and Hawking-Moss solutions.
In fact, they can be viewed as, in a sense, interpolating between
these two limiting cases.

These oscillating bounces exist on a space that is topologically a
Euclidean four-sphere.  They are O(4)-symmetric, so that the scalar
field depends only on the parameter $\xi$, which is proportional to
the azimuthal angle.  At $k \ge 1$ values of $\xi$ the scalar field
takes on its value at the top of the potential barrier.  These values
correspond to $k$ three-spheres that can be viewed as dividing the
four-sphere into two end caps and $k-1$ bands.  Typically, in each of
the end cap regions the field approaches close to, but does not quite
reach, one of the two possible vacuum values.  In the intermediate
band regions the field is alternately on the true or false vacuum side
of the barrier, although it does not generally wander very far from
the top of the barrier, with this deviation decreasing as $k$
increases.

For $k=1$, the solution is simply the Coleman-De Luccia
bounce.  The $k=2$ solution for which the field in both end caps is in
the true vacuum region can be viewed as a two-bounce solution.
However, the solutions with higher values of $k$ are not in any sense
multi-bounce solutions.  For any fixed value of the parameters of the
theory, there are only a finite number (possibly zero) of these
solutions.  When solutions exist, they have values of $k$ taking on
all integer values from 1 to some $k_{\rm max}$, with possibly several
distinct solutions for the same value of $k$.  The determination of
precisely how many solutions exist is closely related to the older
problem of determining when a potential admits a Coleman-De Luccia
bounce.

In a rough sense, the controlling quantity here is the ratio of the
second derivative of the potential to the value of $H^2$.  This is
physically plausible, since for ``typical'' potentials $V''$ is
related to the characteristic mass scale, which in turn determines the
characteristic size of a classical solution.  If this last quantity
were too large, one might imagine that the bounce would not ``fit'' on
the four-sphere.
Early discussions focused on $\beta \equiv |V''(\phi_{\rm top})|/H^2$,
and argued that there was no Coleman-De Luccia bounce if $\beta < 4$.
We have shown that for a broad class potentials $\beta$ is indeed the
controlling parameter and that $k_{\rm max}= N$, where $N$ is the
largest integer such that $N(N+3) < \beta$.  As $\beta$ is
increased through one of these critical values, two (possibly
equivalent) $k=N$ solutions appear.  Precisely at the critical value,
the new solutions are identical to the Hawking-Moss solution, and so
can be thought of as being ``all wall''.  (As described in Sec.~IV,
matters can be slightly more complicated if $k$ is an even number.)

Not all potentials give this behavior, as is evident from our results for the
case where the fourth derivative of the potential is negative at the
top of the barrier.  Here, the critical values of $\beta$ give upper,
rather than lower, limits for the existence of small amplitude bounce
solutions.  The situation is somewhat clarified by considering
potentials where the barrier is atypically flat.  For these, the
parameter whose value determines the first appearance of a bounce is
not the second derivative at the top of the barrier.  Instead, one can
use an averaged value of $|V''|/H^2$ (whose precise definition may
vary from case to case) to define a quantity $\gamma$.  As $\gamma$ is
increased through a $k$-dependent critical value $\gamma_1$, a pair of
$k$-solutions appears.  These coincide, and are nontrivial, when they
first appear.  In the simplest case, one of these will persist for all
$\gamma >\gamma_1$, while the other survives only for a finite range
$\gamma_1 < \gamma < \gamma_2$, with the parameters that give
$\gamma=\gamma_2$ also giving $\beta = k(k+3)$.  When this critical
point is reached, the latter solution and its $\xi$-reversed image
annihilate and merge into the Hawking-Moss solution.  More complicated
patterns are also possible, but must satisfy the constraint that
solutions only appear and disappear in pairs.  To satisfy this
requirement there must be at least one solution that survives for
arbitrarily large $\gamma$, and a second solution that either merges
with the Hawking-Moss at a finite value of $\gamma$ or, as with
our toy model, also persists for arbitrarily large $\gamma$.

\subsection{Robustness of our results}

To simplify our analysis, we have imposed two restrictions.   First,
we have only considered solutions with O(4) symmetry.  Second, we
have assumed that the potential is such that we can use the fixed
background approximation.  We must now ask how robust our results are,
and how they might change were these restrictions relaxed.

The restriction to O(4)-symmetric configurations is the common
practice in flat space treatments of vacuum tunneling.  For
single-field potentials satisfying a rather plausible set of
conditions, it can even be shown that the lowest action bounce has
O(4) symmetry.  Indeed, although there are nonsymmetric approximate
stationary points corresponding to several widely separated bounces,
it may well be that most theories admit no bounce solutions
without O(4) symmetry.

This situation is no longer the case when gravitational corrections
are included, largely because the Euclidean space is now a compact
manifold.  To see this, consider the regime where the true vacuum
bubble of the Coleman-De Luccia bounce has a radius much less than
$H^{-1}$.  Now arrange several such bubbles in a maximally symmetric
fashion around the four-sphere (e.g., at the six vertices of a
five-dimensional hypertetrahedron).  It is clear that some mild local
smoothing of the configuration will give a stationary point of the
action, and hence a bounce solution without O(4) symmetry.  The
solution will, however, be invariant under a discrete subgroup of
O(5).  It is quite possible that there might be additional solutions,
with oscillatory behavior, with the same discrete symmetry.  However,
there no reason to expect more than a finite number of these.
Further, we would expect that the actual number of such solutions
would have a dependence on the parameters of the theory that is similar
to that found for the O(4)-symmetric solutions.

A more significant limitation on our analysis is the use of the fixed
background approximation, which is clearly inapplicable
to many potentials of interest.  Nevertheless, we expect most if not
all of the qualitative, and indeed many of the quantitative, features
of our results to persist.  Thus, for solutions that begin as small
deviations from the Hawking-Moss solution when $\beta$ reaches a
critical value (either from above or below), the value of $\beta_{\rm cr}$
will be exactly as in our analysis, provided that $H$ is calculated
using $V(\phi_{\rm top})$.

A second case to consider is the toy model of Sec.~VI.  Let us assume
that the linear falloff of the potential is terminated at some finite
value of $|\phi|$ in such a way that $V(\phi)$ remains everywhere
positive.  Further, let $\gamma$ still be defined using the
value of the potential at the top of the barrier.  The fact that
$V(\phi)$ is actually lower than this in some regions would have the
effect of decreasing the effective value of $H$; from
Eq.~(\ref{EucFriedmann}), we see that the nonzero $\dot\phi$ acts in the
same direction.  The Euclidean space will therefore be larger, making
it easier to fit in a bounce solution.  As a result, we expect a
reduction in the values of $\gamma$ at which solutions for a given $k$
will first appear, and thus an even larger range of parameters that
allow a bounce with an absolutely flat potential.

\subsection{Physical interpretation}

We now turn to the physical interpretation of these solutions.  As
with the Coleman-De Luccia and Hawking-Moss bounces, the hypersurface
passing through $\xi=0$ and $\xi=\xi_{\rm max}$ gives the initial data
for the classical evolution after the vacuum transition governed by
the bounce.  This hypersurface is a three-sphere of
radius $\sim H^{-1}$, apparently implying that the transition takes place
at the ``waist'' of the de Sitter hyperboloid.  This is generally 
understood to be an artifact of the formalism, with the physical 
results instead being adapted to a single horizon volume.

Formally, the classical evolution can be obtained by an analytic
continuation of the bounce solution.  In doing so, particular care
must be taken at the points $\xi=0$ and $\xi=\xi_{\rm max}$ of the
Euclidean solution.  After rotation to a Lorentzian signature these
become the lightlike boundaries of two antipodal regions on the de
Sitter hyperboloid~\cite{Guth:1982pn}; this generalizes the manner in
which the origin of the flat space Euclidean solution becomes a
lightcone after the Wick rotation to Minkowski space.

In the case of the Coleman-De Luccia bounce, this procedure yields a
Lorentzian spacetime in which portions of two de Sitter spaces,
corresponding to the two vacua, are separated by a well-defined bubble
wall.  When the initial bubble size is small, the bounce can be viewed
as having an initial state hypersurface corresponding to the false
vacuum (see Fig.~2), and the vacuum transition is most naturally
interpreted as quantum mechanical tunneling.  When the initial bubble
size is comparable to $H^{-1}$, the initial state is not evident in
the bounce, but is instead determined only by which vacuum action is
subtracted in the calculation of $B$.  The transition thus has a more
thermal character, reflecting the existence of a nonzero de Sitter
temperature $T_{\rm dS}$.

The Hawking-Moss solution is even more clearly thermal in character.
Here the initial data for the classical evolution is a homogeneous
region with the field balanced at the top of the potential barrier.
Although, strictly speaking, the classical Lorentzian evolution would
leave the scalar field at the top of the barrier forever, this
solution is unstable against small fluctuations and so will break up,
in a stochastic fashion, into regions that evolve toward one or the
other of the two vacua.

The oscillating bounces yield a hybrid of these two cases.  The end
cap regions near $\xi=0$ and $\xi=\xi_{\rm max}$ evolve into two
vacuum regions bounded by well-defined walls; these vacua can be the
same or different, and have no definite relation with the state before
the vacuum transition.  In the intermediate region, the scalar field
profile oscillates about the top of the barrier\footnote{An exception
is the $k=2$ ``true-to-true'' solution, in which the intermediate
region is approximately false vacuum at nucleation.}.  With initial
conditions given precisely by the bounce solution, these oscillations
would be preserved under the classical evolution, giving a large and
exponentially expanding region in which the field is near the top of
the barrier.  However, just as with the Hawking-Moss solution, there
will be instabilities against small fluctuations.  We expect these to
lead to a breakup into a stochastic mixture of regions evolving toward
the true and false vacua, separated by relatively narrow transition
regions.  (We differ here from the scenario advocated in
Ref.~\cite{Banks:2002nm}.)

The relative importance of these three types of solutions depends on
the parameters of the theory.  For ``typical'' potentials, such as
that studied in Sec.~V, the various regimes can be characterized by
the value of $\beta$.  (If the potential is atypical, e.g., by being
unusually flat at the top of the barrier or by having two very
different mass scales, the regimes will be defined by a more complex
combination of parameters, but the discussion would be analogous.)
When $\beta \gg 1$, there will be a Coleman-De Luccia bounce, a
Hawking-Moss solution, and many oscillating bounces.  However, the
actions of the Hawking-Moss solutions and the $k>2$ oscillating
bounces will be many times greater that of the Coleman-De Luccia
bounce, which will clearly dominate.  This is a regime of quantum
tunneling transitions followed by deterministic classical evolution.
At the other extreme is the case $\beta \lesssim 1$, where only the
Hawking-Moss solution is the only bounce.  This is a regime of thermal
transitions followed by stochastic evolution.
Intermediate between these is a regime of moderate $\beta$ that admits
the Hawking-Moss, one (or more) Coleman-De Luccia, and several 
oscillating bounces.  It is in this transitional regime that the 
oscillating bounces are most likely to play a role.

Finally, we should comment on the question of negative eigenmodes.  In
the case of flat space quantum tunneling, the existence of a single
negative mode (or, possibly, a number equal to one mod four) was an
essential ingredient to the path integral derivation of $\Gamma$.
Indeed, it can be shown~\cite{Coleman:1987rm} that for a wide
class of theories in flat space the bounce of lowest action has only
one negative mode.  The situation in curved space, and in particular
for the oscillating bounces, needs further clarification, both as to
how these affect the contribution to the vacuum transition rate and
to their role in the stochastic breakup of the intermediate regime
after the transition.  We plan to return to this subject in a later
publication.

\acknowledgments

We would like to thank Andrei Linde for helpful conversations.
This work was supported in part by the U.S. Department of Energy.


\end{document}